\documentclass[letterpaper]{aa}
\usepackage{graphics}
\usepackage{txfonts}
\usepackage{natbib}

\def\xmm{{\em XMM-Newton}\ }
\def\suzaku{{\em Suzaku}\ }
\def\chandra{{\em Chandra}\ }

\begin{document}

\title
{
An absorption origin for the X-ray spectral variability of MCG--6-30-15
}
\subtitle{}

\titlerunning{The X-ray spectral variability of MCG--6-30-15}

\author
{
L.\ Miller\inst{1} \and 
T.\ J.\ Turner\inst{2,3} \and 
J.\ N.\ Reeves\inst{4} 
}

\authorrunning{L.\ Miller et al.\ }

\institute{
Dept. of Physics, University of Oxford, 
Denys Wilkinson Building, Keble Road, Oxford OX1 3RH, U.K.
\and
Dept. of Physics, University 
of Maryland Baltimore County, 1000 Hilltop Circle, Baltimore, MD 21250, U.S.A.
\and 
Astrophysics Science Division,   
NASA/GSFC, Greenbelt, MD 20771, U.S.A.
\and 
Astrophysics Group, School of Physical and Geographical Sciences, Keele 
University, Keele, Staffordshire ST5 5BG, U.K.
}

\date{Received / Accepted}

\abstract
{
The Seyfert\,I galaxy MCG$-6$-30-15 shows one of the best examples of
a broad ``red wing'' of emission in its X-ray spectrum at energies 
$2 < E < 6.4$\,keV, commonly interpreted as being caused by
relativistically-blurred reflection close to the event horizon of
the black hole.
}
{
We aim to test an alternative model in which absorption 
creates the observed spectral shape, explains the puzzling
lack of variability of the red wing and reduces the high reflection albedo,
substantially greater than unity, that is otherwise inferred at energies
$E > 20$\,keV.
}  
{
{We compiled all the available long-exposure, high-quality data for MCG$-6$-30-15:
522\,ks of \chandra {\sc hetgs}, 
282\,ks of \xmm {\sc pn}/{\sc rgs} and 
253\,ks of \suzaku {\sc xis}/{\sc pin} data.
This is the first analysis of this full dataset.  We investigated
the spectral variability on timescales $>20$\,ks using principal components
analysis and fitted spectral models to ``flux state'' and mean spectra
over the energy range $0.5-45$\,keV (depending on detector).
The absorber model was based on the zones previously identified
in the high-resolution grating data.  Joint fits were carried out to any
data that were simultaneous.}
}
{
Multiple absorbing zones covering
a wide range of ionisation are required by the grating data, including a
highly ionised outflowing zone. A variable partial-covering zone
plus absorbed low-ionisation reflection, distant from the source, 
provides a complete description of the variable
X-ray spectrum. A single model fits all the data.
We conclude that these zones are 
responsible for the red wing, its apparent lack of variability, the absorption
structure around the Fe\,K$\alpha$ line, the soft-band ``excess'' and the 
high flux seen in the hard X-ray band. 
A relativistically-blurred Fe line is not required in this model.
We suggest the partial covering zone
is a clumpy wind from the accretion disk.
} 
{}

\keywords{accretion; galaxies: active}

\maketitle

\section{Introduction}
The galaxy MCG$-6$-30-15 is a well-known Seyfert type I at redshift $z=0.00775$,
and was one of the first to show evidence in ASCA observations for a broad wing 
of emission extending to lower energies from the 6.4\,keV\,Fe\,K$\alpha$ line
\citep{tanaka95}.  A common interpretation of this emission is that it arises
as reflection from the inner regions of the accretion disc, where the photon
energy is smeared to low energies by relativistic effects 
\citep[e.g.][]{tanaka95,iwasawa96}.  The existence of the ``red wing'' has been
subsequently confirmed by more recent observations with \xmm and \suzaku
(\citealt{wilms}, \citealt{fabian02}, \citealt{vaughanfabian04}, 
\citealt{reynolds04}, \citealt{miniutti07}), but 
investigation of models for the physical origin of this component have concentrated
on the ``blurred reflection'' hypothesis (see \citealt{brenneman06} for a
summary of the history of analysis of the X-ray spectrum).

However, it has long been known that absorption can conspire to
produce continuum shapes similar to those observed in the
AGN with red wings (e.g. in NGC\,3516, \citealt{turnerea05}
and NGC\,3783, \citealt{reeves04}), and
MCG$-6$-30-15 itself is known to show strong absorption lines
indicating the presence of absorption by zones of gas with a very
broad range of ionisation.  In the Fe\,K$\alpha$ region both
\citet{young05} in high-resolution \chandra {\sc heg} data and
\citet{miniutti07} in \suzaku {\sc xis} data detected lines at 6.7 and
7.0\,keV, most likely identified as \ion{Fe}{xxv} and \ion{Fe}{xxvi}
with an outflow velocity $\sim 1800$\,km\,s$^{-1}$.  This
identification of a highly ionised outflowing zone is substantiated by
the detection of matching lines from \ion{Si}{xiv} and \ion{S}{xvi} at
2.0 and 2.6\,keV in the \chandra {\sc meg} data \citep{young05}.  
High-resolution data at softer energies also reveal lines from a
broad range of ionisation: \citet{lee01} showed that at least two zones
of gas with ionisation parameter $\log\xi \simeq 0.7$ and 2.5 were required
to explain the lines detected in the \chandra {\sc meg} data, and these
lines were further  confirmed by detections in \xmm {\sc rgs} data
\citep{turner03, turner04}.  \citet{lee01} also interpreted the 
strong edge observed at 0.7\,keV as being an edge from \ion{Fe}{i}, possibly
arising in dust grains.  To date there has been little published evaluation
of the possible effect of such zones on the observed continuum from this source,
although some analyses have included some of the known absorption zones
\citep[e.g.][]{brenneman06}:
in this paper we shall investigate absorption-dominated models that both explain the
continuum shapes and the absorption lines that are observed.

A further motivation for this work is that 
it has been recognised for some time that the red wing component is surprisingly
constant in amplitude \citep{iwasawa96,vaughanfabian04}
(although \citealt{ponti04} did find evidence for additional
variability around 5\,keV in the 2000 \xmm observation).
If the red wing
emission is indeed reflected emission from the accretion disk we would
expect its amplitude to vary in phase with the primary continuum, counter
to what is found in most observations: the only occasion on which it
has been inferred that this expected property did occur was when the source
was in its lowest flux state, as observed with \xmm in 2000 
\citep{reynolds04}.  
The general lack of red wing
variability has led to the development of a ``light-bending''
model in which the observed primary continuum variations are caused by variations
in height of the illuminating continuum coupled with distortions of photon
geodesics near the black hole, rather than by rest-frame intensity variations,
so that the reflected emission appears more constant than the primary continuum
seen by the observer \citep{fabianvaughan03, miniutti03, miniutti04}.  
The same model has also been
invoked to explain the high flux observed at $E>20$\,keV in \suzaku {\sc pin}
data as being the Compton reflection ``hump'' enhanced in amplitude by the
same light-bending effects \citep{miniutti07}.  
Other models that seek to explain the constant
red wing amplitude \citep[e.g.][]{merloni06, nayakshin, zycki04} 
do not explicitly produce an
explanation for the high energy excess, and some are unlikely to apply to
MCG$-6$-30-15 \citep{zycki01}
and on long timescales \citep[][hereafter M07]{miller07}.
However, light-bending is not the only way
to produce a constant red wing and a high energy excess.  The constancy of the
component could be explained if the reflection were distant from the primary
source, so that light travel time smooths out any illumination variations,
or models with a variable covering fraction of absorption can produce a
similar effect (see the discussion in M07
and \citealt{turner07}, hereafter T07).
Furthermore, the high energy excess may be explained as either being a combination
of reflection with no light-bending enhancement and a contribution from a
highly absorbed (N$_{\rm H} \ga 10^{23}$\,cm$^{-2}$)
continuum component: or indeed perhaps even an absorbed component alone.
The existence of a high-energy excess on its own does not require light-bending
or relativistic blurring of reflected emission.

Thus our aim in this paper is to investigate the
extent to which the full X-ray spectrum of MCG$-6$-30-15 may be explained by the
effect of absorption of the intrinsic emission.  We are aided
in this by two key advances since previous analyses.  First, there now exists 
a significant body of high quality data, from three independent observatories, \chandra,
\xmm and \suzaku.  The \suzaku data include low-resolution spectral measurements
up to $\sim 45 $\,keV, and in this paper we analyse, for the first time, the entire
set of CCD-resolution data that is available for this source, and we test physical 
models against the full energy range available, $\sim 0.5-45$\,keV.  We also test
the models for consistency with the high-resolution \chandra {\sc hetgs} and \xmm
{\sc rgs} data.
 
The second advance
is to make as much use as possible of the spectral variability exhibited by this
(and other) AGN:  the spectrum appears to change shape in a systematic way as the
total X-ray flux of the source varies, implying that the emission we see is made up
of a number of components whose individual spectra differ.  We use a method
of principal components analysis, described below, that retains the full spectral
information content of the data, to first set the basic parameters of models describing
each of those components, and then we test the resulting model against the actual 
data.

\section{Analysis methods}
\subsection{Principal components analysis}\label{pca}
The starting point for the physical models developed in this paper is
a model-free principal components analysis (hereafter PCA)
of the spectral variations
of MCG$-6$-30-15.  Such an analysis was first applied to this source
with low spectral resolution by \citet{vaughanfabian04}.  In this paper
we use the method developed and described by M07, which uses
singular value decomposition to achieve a principal components decomposition
of the spectral variations whilst retaining the full spectral resolution
of \xmm {\sc pn} and \suzaku {\sc xis} data.
In the case of the analysis of Mrk\,766 (M07) this provided significant
additional information, as it revealed the presence of absorption edge and line
features associated with a hard spectral component, and it provided
model-independent evidence for the presence of both ionised absorption associated
with the principal varying component and also ionised Fe emission.  We shall see
below that similar features appear in MCG$-6$-30-15.

The optimal-resolution 
PCA procedure adopted is identical to that of M07 and we refer to
that paper for full details of the method.  
The random errors in the 
principal components arising
from photon shot noise are estimated by a Monte Carlo simulation, in which the PCA is
repeated for synthetic spectra that are perturbed by random noise about the true
spectra.  The resulting errors are correlated between the principal components but do
allow some estimation of the effect of shot noise.

The basic output from the PCA is a set of eigenvectors, each one representing the 
spectrum of an additive mode of variation.  The generic varying power-law, that is a
ubiquitous feature of AGN X-ray spectra, appears as ``eigenvector one'' in this analysis.
In addition it was found in Mrk\,766 that the bulk of the variation could be
described by a single eigenvector superimposed on a quasi-constant component of emission,
which in that paper was named the ``offset component''. 
But there is not a unique
association of physical components with the eigenvectors and offset components.  For
example, there could in the source be components with exactly the spectra of eigenvector
one (a varying absorbed power-law) and the offset component (which could
be reflected emission from an extended region where light-travel-time effects dampen any
intrinsic variation in the illumination). Alternatively
there could instead be two variable additive components
whose variations are correlated, such that eigenvector one represents the net overall
effect of their joint variation.  A possible physical model that achieves the latter is
if spectral variations are caused by variations in the covering fraction of an
absorber passing across an extended source - in this case the offset component represents the
view of the source when it is most covered, the highest flux state represents the source
when it is most uncovered, and eigenvector one represents the difference between these
states.  These interpretations can produce identical PCA results and are indistinguishable:
we must rely on physical models to attempt to discern which interpretation is closest to
reality.  In practice, it may be that both a quasi-constant reflection
component and a variable covering-fraction
absorber are present, a hybrid model also mentioned by M07 and T07.

Such an analysis effectively assumes that the spectral variations may be described
by additive spectral components.  In the case of Mrk\,766 it was found that although
the chief variations could be modelled in this way, there were still significant
non-additive variations which were found to be caused by absorption variations
(M07, T07).  In this parallel analysis of MCG$-6$-30-15 we shall
first investigate the PCA, and fit models to the principal components, and then
also explore directly fitting to the data the models that arise from the PCA.
We shall not, however, explore temporal absorption variations, although we shall see
that there is evidence for these also in MCG$-6$-30-15.

\subsection{Data grouping and spectral fitting}
The principal statistical tool used here will be goodness-of-fit
testing using the binned $\chi^2$ statistic.  
As in M07, we adopt an optimum spectral binning of the
data, with spectral bins whose width in energy are equal to half the
energy-dependent instrumental FWHM.  Any spectral features arising
in finer binning cannot be intrinsic to the source and we should not
fit models with binning finer than this.  It is commonplace in the 
literature on X-ray observations for goodness-of-fit statistics to
be quoted with much finer binning than the instrumental resolution: 
such reported values can be
misleading, in that the reduced $\chi^2$ in such cases may appear
low, but the goodness-of-fit probability might in fact be rather poor.
The sensitivity to departures
of the model from the data is significantly reduced with binning that is
too fine.
To illustrate this here with a simplified example, suppose data is divided into $n$ bins,
fit by a model with $p$ free parameters 
that yields a goodness-of-fit $\chi^2_n = n - p + \Delta\chi^2$,
where $\Delta\chi^2$ is the excess $\chi^2$ over the expectation value $n-p$
for a well-fitting model.  If the data
are then more finely divided into $m$ bins, $m>n$, 
with the data errors increasing as the 
inverse square root of the bin width, then if the model does not
have any variations on this finer scale we can use the properties of the
non-central $\chi^2$ distribution to infer an expected new 
goodness-of-fit, $\chi^2_m \simeq m - p + \Delta\chi^2$ where $\Delta\chi^2$
has the same value as in the case of $n$ bins.
For example, for
$\chi^2_n=400$, $n=300$, we would obtain $\chi^2_m=3100$ for $m=3000$
(adopting the simplification $p=0$).
The first case has a reduced $\chi^2=1.33$ and the model would be rejected
at significance level $p=10^{-4}$.  The second case has reduced $\chi^2=1.033$
and the model would only be rejected at $p=0.1$.

In this paper we adopt the optimum energy binning for maximum sensitivity, 
and hence the goodness-of-fit statistics will indicate worse fits than if we had adopted
finer energy binning.  Where relevant, we shall compare goodness-of-fit with
previous published results for MCG$-6$-30-15 by calculating $\chi^2$ with
similar binning to that adopted by other authors.

\subsection{Systematic uncertainties in data and models}

Photon counts are 
everywhere sufficiently high for the approximation that the
error in each bin has a normal distribution to be valid.
However,
the datasets being analysed comprise long observations on this
bright AGN, and in fact the goodness-of-fit is no longer 
dominated entirely by photon statistics, especially at the
low energy end of the spectra.  

Systematic errors in the data exist through uncertainty in the 
calibration.  These uncertainties have not yet been fully quantified
for \suzaku, but for \xmm it seems that there can be energy-dependent
discrepancies of 5-10\,percent between {\sc pn} and {\sc mos} instruments,
depending on the spectrum of the source, and discrepancies of
up to 20 percent in the \xmm {\sc pn} v. \suzaku {\sc xis} cross-calibration
\citep{stuhlinger}.

There are also systematic uncertainties in the models, in the sense
that, in order to keep the number of free parameters to a minimum,
we inevitably fit simplified models to what in reality must be
a complex set of emission and absorption processes.  For example,
it is commonplace to use the {\sc reflion} models of \citet{rossfabian}
to model reflection spectra.  However, those models assume 
a constant density slab at normal inclination, whereas in practice
there must be a complex geometry.  Emission-line fluxes in particular
will be orientation dependent \citep[e.g.][]{george91, zycki94}, and 
it has been suggested that a disk atmosphere in hydrostatic pressure
equilibrium would result in a quite different spectrum 
\citep{donenayakshin}.  Such systematic ``model error'' is
difficult to quantify.

There is no rigorous way of taking such systematic errors into account
in the goodness-of-fit testing.  Within {\sc xspec} \citep{arnaud}
it is possible
to treat the systematic error as being a constant fractional error
that is added in quadrature with the random error.  In this paper we
adopt as standard a fractional systematic error of 3\,percent, a value
that is around the lowest systematic uncertainty we might expect given
the cross-instrument comparisons that have been made to date.  Where
relevant, when comparing our results with previous results in the literature,
we shall quote goodness-of-fit statistics assuming zero systematic error.

\section{The data}

\subsection{XMM-Newton data}
The \xmm dataset comprises two observations made in 2000,
previously described and analysed by \citet{wilms}, \citet{reynolds04}
and \citet{ponti04},
and three closely-spaced observations made in 2001,
previously described and analysed by 
\citet{fabian02}, \citet{vaughanfabian04} and \citet{brenneman06}.
The observation dates, IDs, duration of the observations and on-source
exposure times after screening are given
in Table\,\ref{tabledata}.

\begin{table}
\caption
{Datasets used in this analysis, giving observation date, ID,
observation duration and on-source exposure time 
({\sc pn} times are given for \xmm, {\sc xis} times are given for \suzaku).}
\label{tabledata}
\begin{tabular}{lrlrr}
\hline\hline
Observatory & \multicolumn{1}{c}{date} & \multicolumn{1}{c}{ID} 
& dur & exp \\
 & & & /ks & /ks \\
\hline
\xmm &  11 Jul 2000 & 0111570101 & 43.2 & 41.1 \\
\xmm & 11-12 Jul 2000 &  0111570201 & 46.0 & 32.3 \\
\xmm & 31 Jul 2001 & 0029740101 & 79.5 & 57.7 \\
     & - 1 Aug 2001 &  & &  \\
\xmm & 2-3 Aug 2001 & 0029740701 & 125.9 & 69.0 \\
\xmm & 4-5 Aug 2001 & 0029740801 & 125.0 & 81.5 \\
\suzaku & 9-14 Jan 2006  & 700007010 & 354.2 & 103.5 \\
\suzaku & 23-26 Jan 2006 & 700007020 & 215.3 & 72.1 \\
\suzaku & 27-30 Jan 2006 & 700007030 & 208.3 & 77.4 \\
\chandra & 19-27 May 2004 & 4759-4762 & 530.0 & 521.8 \\
\hline
\end{tabular}
\end{table}

In this analysis we use data from the {\sc epic pn} CCD detector 
\citep{struder} in the energy range $0.4-9.8$\,keV. Data from the 
Metal Oxide Semi-Conductor ({\sc mos}) CCDs  were not used as they suffer significant 
photon pileup and inferior signal-to-noise.  
All {\sc pn} observations utilised the medium filter and the small window mode. 
Data were processed using {\sc sas v7.0} 
using standard criteria with instrument patterns 0--4 and removing periods 
of high background (where the rate in the background cell exceeded 
0.15\,count\,s$^{-1}$).  Source data were extracted 
from a circular cell of radius $50''$ centred on the source, and 
background data were taken from a source-free region of approximately the same size 
within the same {\sc pn} chip. 
The total 2000 and 2001 {\sc pn} exposures were 73\,ks and 208\,ks, respectively. 
The typical deadtime correction was a factor 0.7.
During 2000 
the  mean {\sc pn} count rate was $\sim 3.709$ $\pm0.010$\,count\,s$^{-1}$  
and during 2001 
 $\sim 4.908$ $\pm0.005$\,count\,s$^{-1}$ in the 2--10 keV band. 
The mean background level in the screened data was $< 1$\%  of the 
mean source rate in this band. 

We also used higher resolution spectra from the Reflection Grating Spectrometer
({\sc rgs}, \citealt{denherder01}) which 
were taken from the pipeline processing and were coadded for each {\sc rgs} 
grating to yield two spectra for  
each of the 2000, 2001 epochs. As the individual exposures were 
taken just days apart and the source was centred the same on the detector for each 
observation, 
the responses were indistinguishable for the parts that were coadded and this 
summation did not result in any significant loss of effective resolution within the data. 
The total {\sc rgs 1} exposure was 108\,ks for 2000 and 331\,ks for 2001; 
the {\sc rgs 2} exposure was 105\,ks 
for 2000,  323\,ks  for 2001.   During 2000 first-order 
{\sc rgs} data yielded 0.71$\pm 0.003$\,count\,s$^{-1}$ 
and $0.64 \pm 0.03$\,count\,s$^{-1}$ for the summed 
{\sc rgs 1} and {\sc rgs 2} data respectively, and 
during 2001, 0.89$\pm 0.002$\,count\,s$^{-1}$  
and 0.90$\pm 0.02$\,count\,s$^{-1}$ for {\sc rgs 1} and {\sc rgs 2}, respectively.
The {\sc rgs} background level was $\sim 11\%$ of the total count rate.     

\subsection{\suzaku data}
The \suzaku dataset comprises three observations made in 2006 as
tabulated and previously described and analysed by \citet{miniutti07}.
Because of the much shorter duration and the changing instrumental
response we do not use a shorter observation made on 2005 Aug 17.  The
observation dates, IDs, total duration of the {\sc xis} observations
and on-source exposure times after screening are given in
Table\,\ref{tabledata} ({\sc xis 0} typically had a slightly lower
on-source exposure time than the other detectors).  We used {\sc xis}
and {\sc hxd pin} events from v2.0.6.13 of the \suzaku pipeline. Both
the {\sc xis} and {\sc hxd pin} data were reduced using v6.3.2 of {\sc
HEAsoft} and screened with {\sc xselect} to exclude data during and
within 436 seconds of entry/exit from the South Atlantic Anomaly
(SAA).  Additionally we excluded data with an Earth elevation angle
less than 5$^\circ$ and Earth day-time elevation angles less than
20$^\circ$.  A cut-off rigidity (COR) of $>6$\,GeV was applied to
lower particle background.  The source was observed at the nominal
centre position for the {\sc xis} for all observations. The {\sc
xis-fi} CCDs were in $3 \times 3$ and $5 \times 5$ editmodes.  Data
from the back-illuminated {\sc xis} 1 detector were not used because
its effective area is lower than that of the front-illuminated XIS
0/2/3 at 6 keV, while its background rate is larger at high energies.
For the {\sc xis} CCDs we selected good events with grades 0,2,3,4,
and 6 and removed hot and flickering pixels using the {\sc sisclean}
script.

The {\sc xis} products were extracted from circular regions of
2.9\arcmin radius while background spectra were extracted from a
region of the same size offset from the source (and avoiding the chip
corners with the calibration sources).  The response matrix (rmfs) and
ancillary response (arf) files were then created using the tasks {\sc
xisrmfgen} and {\sc xissimarfgen}, respectively. {\sc xissimarfgen}
accounts for the hydrocarbon contamination on the optical blocking
filter.  The background was $<1\%$ of the total {\sc xis} count rate
in the full {\sc xis} band for each CCD. The {\sc xis} 0,2,3 were
combined to produce a single {\sc xis-fi} spectrum, along with the
response files with the appropriate (1/3) weighting.

The {\sc pin} background events file was provided by the {\sc hxd}
instrument team, used in conjunction with the source events file to
create a common good time interval set applicable to both the source
and background.  The background events file was generated using ten
times the actual background count rate, so we increased the effective
exposure time of the background spectra by a factor of 10.  We found
the deadtime correction factor using the {\sc hxddtcor} task with the
extracted source spectra and the unfiltered source events files.  The
total on-source exposure time, after screening, was 292\,ks, somewhat
longer than for the {\sc xis}.  The contribution of the cosmic X-ray
background \citep{boldt87,gruber99} was taken into account as
described later.  The source comprised 26\% of the total counts,
clearly above the $1\sigma$ 3.2\% {\sc hxd pin} background systematic
level. The response file ae\_hxd\_pinxinome1\_20070914.rsp provided by
the instrument team was used in all the spectral fits.  In the
subsequent analysis, the {\sc pin} flux has been decreased by a factor
1.09 to be consistent for observations at the {\sc xis} pointing
position with the cross-calibration study of \citet{koyama}, revised
for the \suzaku revision\,2 analysis pipeline.\footnote{The cross
normalisation of the {\sc hxd/pin} vs. {\sc xis-fi} has been
determined to be in the range 1.06--1.09 for {\sc xis} nominal
pointing from observations of the Crab (see
ftp://legacy.gsfc.nasa.gov/suzaku/doc/xrt/suzakumemo-2007-11.pdf).}

\subsection{\chandra {\sc hetgs} data}
{\it Chandra} {\sc hetgs} exposures for MCG$-6$-30-15 were taken
during May 2004 yielding 522\,ks of good data
(OBSIDs 4759-62) as detailed by \citet{young05}
(Table\,\ref{tabledata} gives the summed duration of each of the 
four OBSIDs). An earlier
observation made in 2000 has been analysed by \citet{lee01} but is not
included here owing to its much shorter duration.  Data were reduced
using {\sc ciao} v3.4 and {\sc caldb} v3.4.0 and following standard
procedures for extraction of HETG spectra, except that we
used a narrower extraction strip than the {\sc tgextract} default.
This reason for the default processing cut-off is that the overlap of
the MEG and HEG strips depends on the extraction strip widths, and if
the latter are too large, a larger intersection of the MEG and HEG
strips results, cutting off the HEG data prematurely.  Specifically,
we used \verb+width_factor_hetg=20+ in the tool \verb+tg_create_mask+,
instead of the default value of 35. As the {\sc hetgs} spectra have a
very low signal-to-noise ratio it was necessary to coadd the positive
and negative first-order spectra, and coadd all four OBSIDs, to create
a high quality summed first-order {\sc heg} and {\sc meg} spectra for
fitting.  After such co-addition, we binned the spectra to 4096
channels before grouping.  The summed {\it Chandra} {\sc hetgs}
exposure yielded a 2--7\,keV count rate $\sim 0.1559 \pm
0.0006$\,count\,s$^{-1}$ and $\sim 0.1926 \pm 0.0007$\,count\,s$^{-1}$
in the summed {\sc heg} and {\sc meg} first order spectra,
respectively. The background level in this case is so low as to be
negligible.

\section{Results - the Principal Components}
\subsection{Principal components analysis}
\begin{figure}
\resizebox{0.475\textwidth}{!}{
\rotatebox{-90}{
  \includegraphics{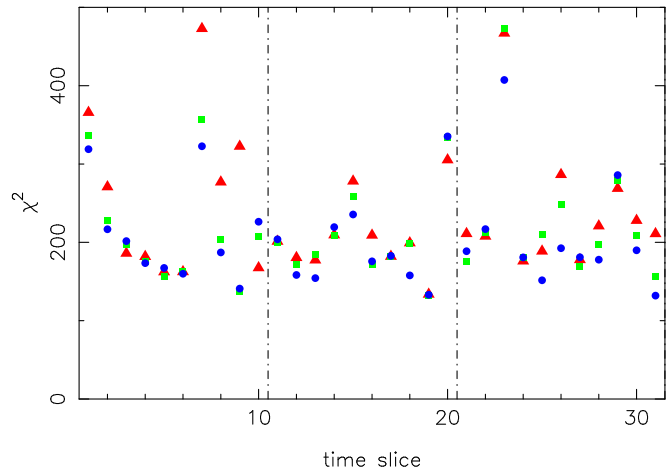}
  }}

\caption{$\chi^2$ values for the full energy range for
each time slice in the \suzaku data,
assuming a constant component plus: a single eigenvector (red triangles);
two eigenvectors (green squares); or three eigenvectors (blue circles).
The number of degrees of freedom is 170-168 respectively.  Dashed vertical
lines indicate the start of a new observation, separated in time from the
previous observation.
}
\label{figchisq}
\end{figure}

We first analyse the \xmm and \suzaku data using PCA, following the
methods described above and in M07.  There are a number of statistical
measures we can employ to test how good the PCA model fits the data,
and how many components are required.  Tables\,\ref{table1} and
\ref{table2} show for the two datasets how much of the observed
variance is accounted for by each component, and how well the data are
described by a source model comprising the offset component plus the
first $n$ eigenvectors, measured by the $\chi^2$ statistic.  The
latter is the mean $\chi^2$ averaged over all timeslices.  It can be
seen that at energies above 2\,keV a single eigenvector is all that is
required in addition to the offset component to adequately describe
the spectral variations. Considering the whole energy range, more
eigenvectors are needed, indicating the effects of variable absorption
as found for Mrk\,766 (M07, T07).  In fact, there are some time slices 
that have anomalously high $\chi^2$ values, as indicated in Fig.\,\ref{figchisq}
which shows for each timeslice in the \suzaku dataset
the value of $\chi^2$ obtained for the full
energy range when allowing an increasing number of components.  Most
time slices are well fit by a single varying component, but there are a 
few time slices, especially in the first observation, where even adopting
3 principal components still leads to a high $\chi^2$ value.  Some of these
``badly fitting'' events are correlated in time with each other, and we suspect
that they are caused by absorption variations, as found for Mrk\,766 (T07).

\begin{table}
\caption{
Statistics of the Principal Components Analysis of the \suzaku data.  
Component 0 indicates
the fit of the mean spectrum alone, subsequent components are eigenvectors
ordered by their eigenvalues (column 2).  Column 3 shows the fractional
variance accounted for by each component, columns 4 and 5 give the 
goodness-of-fit ($\chi^2$ and number of degrees of freedom) for the entire
range and for $2-50$\,keV alone.}
\label{table1}
\begin{tabular}{crrrrrr}
\hline\hline
comp. & eigenvalue & fractional & $\chi^2$ & /dof & $\chi^2$ & /dof \\
 & & variance & \multicolumn{2}{c}{0.4--50 keV} & \multicolumn{2}{c}{2--50 keV}\\
\hline
0 & -       & -     & 4736 & 171 & 1661 & 114 \\
1 & 0.00167 & 0.645 &  235 & 170 &  113 & 113 \\
2 & 0.00052 & 0.201 &  216 & 169 &  108 & 112 \\
3 & 0.00012 & 0.046 &  205 & 168 &  104 & 111 \\
4 & 0.00005 & 0.019 &  191 & 167 &   99 & 110 \\
5 & 0.00002 & 0.009 &  165 & 166 &   93 & 109 \\
\hline
\end{tabular}
\end{table}

\begin{table}
\caption{
Statistics of the Principal Components Analysis of the \xmm data,
with columns as for Table\,\ref{table1} with appropriate energy ranges.
}
\label{table2}
\begin{tabular}{crrrrrr}
\hline\hline
comp. & eigenvalue & fractional & $\chi^2$ & /dof & $\chi^2$ & /dof \\
 & & variance & \multicolumn{2}{c}{0.4--9.8 keV} & \multicolumn{2}{c}{2--9.8 keV}\\
\hline
0 & -       & -     & 22322 & 167 & 3922 & 125 \\
1 & 0.000977 & 0.928 &  953 & 166 &  123 & 124 \\
2 & 0.000016 & 0.015 &  325 & 165 &  105 & 123 \\
3 & 0.000012 & 0.012 &  208 & 164 &   86 & 122 \\
4 & 0.000009 & 0.008 &  161 & 163 &   75 & 121 \\
\hline
\end{tabular}
\end{table}

\begin{figure*}
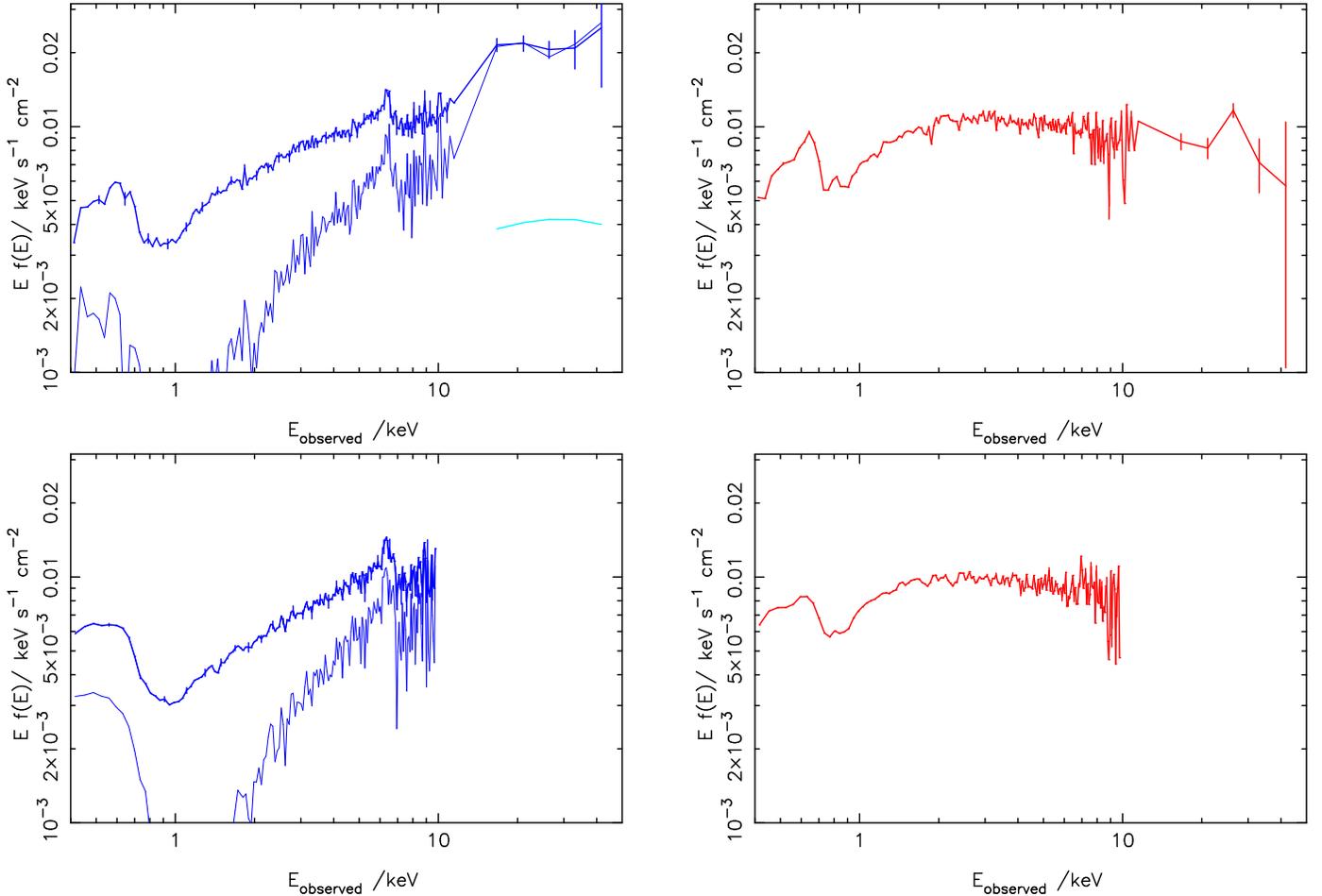

\begin{minipage}{\textwidth}{

\resizebox{0.475\textwidth}{!}{
\rotatebox{-90}{
\includegraphics{9590fig2a.ps}
}}
\hspace*{0.04\textwidth}
\resizebox{0.475\textwidth}{!}{
\rotatebox{-90}{
\includegraphics{9590fig2b.ps}
}}

\resizebox{0.475\textwidth}{!}{
\rotatebox{-90}{
\includegraphics{9590fig2c.ps}
}}
\hspace*{0.04\textwidth}
\resizebox{0.475\textwidth}{!}{
\rotatebox{-90}{
\includegraphics{9590fig2d.ps}
}}

}\end{minipage}
\caption{
Principal component spectra for (top) \suzaku data and
(bottom) \xmm data.  Left-hand panels show limits on the offset
component, right-hand panels show 
eigenvector one, top curve (red), and eigenvector two, lower
curve, falling below the plot boundary (green).
Spectral points in the range $0.4-11$\,keV are from the pn or {\sc xis}
instruments, those in the range $15-45$\,keV are from the 
{\sc pin}.  See text for further details.
}
\label{figpcaspectra}
\end{figure*}

The resulting component spectra are shown in Fig.\,\ref{figpcaspectra}.
The offset component is not uniquely defined, but exists between the limits 
shown:  only a systematic shift between those limits is allowed (more specifically,
the offset component can be shifted by linear combinations of whichever
eigenvectors are allowed to describe the variations).  As well as the {\sc pin} detector
background, a contribution from the 
expected cosmic X-ray background has also been subtracted from the {\sc pin} data, using
the same model and amplitude as \citet{miniutti07}.  This component is also plotted,
so it may be seen that this makes an important but not dominant contribution to
the total measured {\sc pin} flux.

Random uncertainties
on the offset component and eigenvector one are
estimated from a Monte-Carlo realisation as described in section\,\ref{pca}
and are shown on the upper curve only.
Only every fifth error bar is plotted, for
clarity.  Eigenvector one is plotted with an amplitude equal to the mean value 
in each dataset.

For clarity we do not plot further eigenvectors:
although the presence of eigenvector two seems
statistically significant, its amplitude is low and it 
has little effect on the spectral shape of the
offset and first eigenvector components.

\subsection{Spectral models}

\subsubsection{Overview}

As in Mrk\,766, the offset component has a hard continuum shape with a 
soft excess, a weak
line at the energy of low ionisation Fe\,K$\alpha$, an edge and absorption
lines, also detected by \citet{miniutti07} in the total spectrum. 
The {\sc pin} data reveal a high flux in the $15-40$\,keV range as found by
\citet{miniutti07}, although we note that with improved deadtime correction
and background subtraction there is now no indication of any fall-off to
higher energies, consistent with the general lack of cut-offs in AGN spectra
at these energies \citep[e.g.][]{panessa08}.
The first eigenvector has the appearance of an absorbed powerlaw.  The
most basic interpretation of these two components are that the offset component
is a reflection component and eigenvector one is a variable-amplitude
absorbed powerlaw, the variations of which yield the observed spectral variability.
This model has formed the basis for previous analyses of 
the X-ray spectrum of MCG--6-30-15 and other AGN 
(e.g. \citealt{fabian02, fabianvaughan03, vaughanfabian04}) and is the one
we adopt here.  An alternative would be to consider the possible spectral
variations arising from a thermal comptonising plasma 
\citep[e.g.][]{haardt97}, but such models appear to be disfavoured for the
analysis presented here: first, the soft excess is not bright enough
compared with the \citet{haardt97} model; second, in that model it is expected
that the Compton excess should decrease as the $2-10$\,keV flux increases,
counter to what is observed; and finally the PCA presented here provides
strong evidence that the source variability is primarily
well described by the additive variations of a small number of components.

We shall start by fitting basic models to these components.  All fits were
made with {\sc xspec}\,v11 \citep{arnaud}.  Models of ionised absorption
were created using {\sc xstar} \citep{kallman04} which models the absorption
from a spherically-symmetric shell of gas around a central source.
It is particularly important to include the best possible atomic physics
calculations in the absorber models: \citet{kallman04} have shown how
improved calculations lead to bound-free edges that are significantly 
less sharp than would otherwise be obtained, and those authors point out
that this may have a significant effect on the interpretation of the edge
around Fe\,K$\alpha$.  In the models made here, 
fittable parameters were the absorbing column,
the ionisation parameter $\xi$ at the inner face of the absorbing shell
and the redshift.  For simplicity, solar abundances were assumed for all models.
The ionising spectrum was assumed to be a power-law of photon index
$\Gamma=2.2$ between 13.6\,eV and 13.6\,keV: other values of photon index can
produce similar absorption spectra but with a shift in the effective ionisation
parameter.  The gas density was assumed to be $10^{10}$\,cm$^{-3}$ for all
zones except the highest ionisation zone, where a value $10^8$\,cm$^{-3}$
was assumed. 
Although this parameter affects the absorption spectral shape, this again
is largely degenerate with the ionisation parameter, although if the assumed
absorbing shell is sufficiently thick
(as may occur for the combination of high $\xi$ and low density values)
then there can be significant variation in $\xi$ through the zone arising
from inverse-square-law dilution, thus leading to a broader range of ionisation
states than might be the case in a denser, thinner shell of gas.  All the absorption
models are affected by such uncertainties.
We adopt units of erg\,cm\,s$^{-1}$ for $\xi$.

\subsubsection{Joint fit to eigenvector one and offset components}

Qualitatively, the PCA ``eigenvector one'' in both the \suzaku and the \xmm data
has the appearance of a powerlaw affected by ionised absorption.  In reality 
the absorbing zones are complex, and their physical parameters are
unlikely to be unambiguously measurable in data with CCD resolution.  However,
we know {\em a priori} of the existence of ionised absorbing zones from
the previous analyses of \chandra and \xmm high-resolution grating data,
so we can start by seeing whether a model that includes those zones can
explain the CCD spectra we observe.

We first create a model based on
the simplest interpretation of the PCA, namely that the eigenvector one 
represents a variable-amplitude powerlaw, with ionised absorption, and
that the offset component arises from distant reflection, with light
travel-time erasing any reflected amplitude variations.  The primary absorbing
layers that have already been identified in the grating data comprise:
\begin{list}{}{\itemsep=3mm \leftmargin=0mm}
\item {\em Zone 1} with $\log\xi \simeq 2$ \citep{lee01}
\item {\em Zone 2} with $\log\xi \simeq 0.5$ \citep{lee01}
\item {\em Zone 3}, a highly ionised, $\log\xi \ga 3.5$, outflow
at line-of-sight velocity $v \simeq 1800$\,km\,s$^{-1}$ \citep{young05}.
The $6.7$ and $6.97$\,keV lines only appear in the offset component in the
PCA, not on eigenvector one, so for now we assume that this zone is only associated
with the offset component (when fitting to the data later, we allow zone 3
to absorb all components).
This zone
also produces velocity-shifted lines of 2.0\,keV\,\ion{Si}{xiv}\,Ly$\alpha$ 
and 2.62\,keV\,\ion{S}{xvi}\,Ly$\alpha$ which were observed by \citet{young05}
in the \chandra {\sc meg} data.
In this absorption layer the ratio of the 6.7 and 6.97\,keV lines depends on
both the ionisation and on the microturbulent velocity dispersion of the gas,
since the lines are easily saturated, and the relatively high equivalent
width of the lines is most easily achieved by models with line broadening.
Hence the absorption model is broadened by a velocity
dispersion of 500\,km\,s$^{-1}$, consistent with the findings of \citet{young05}
(see section\,\ref{datafitting}), and we fix the ionisation parameter at
$\log\xi = 3.85$ as described later in section\,\ref{models}.
\item {\em Fe\,I edge} at 0.707\,keV.
A further feature apparent in the high-resolution data, but not in data of
CCD resolution, is the complex absorption edge structure at $\sim 0.7$\,keV,
discussed extensively by \citet{branduardi01}, \citet{sako03},
\citet{lee01} and \citet{turner03, turner04}.  Here we adopt the model advocated by the
last three authors, in which the edge primarily arises from neutral Fe which
is possibly in the form of dust grains (see \citealt{ballantyne03dust}
for a discussion of the possible location and origin of the dust).  
Rather than attempting a
complex model of this region, we simply add a single edge at the systemic redshift
whose rest-frame energy corresponds to the 0.707\,keV {\sc l3} Fe\,I edge
\citep{lee01}, with fixed edge optical depth $\tau = 0.4$, as indicated 
by the high-resolution data (in fitting to the grating data later, we allow 
$\tau$ to be a free parameter).
\end{list}

We initially model the offset component as low-ionisation reflection, adopting
the constant density slab model of \citet{ross99} and \citet{rossfabian} and calculated
by the {\sc xspec} extended {\sc reflionx} model made available by those authors.  The 
illumination was assumed to be the same powerlaw for both
eigenvector one and the offset component, and absorbing zones 1 \& 2 and the
dust edge were assumed to cover both components.  
We assume the 6.4\,keV\,Fe\,K$\alpha$ emission line arises
in the low-ionisation reflection, which in the {\sc reflionx} models restricts the
ionisation to be $\xi \la 120$\,erg\,cm\,s$^{-1}$ for photon index $\Gamma \simeq 2.2$.
Better overall $\chi^2$ values may be obtained by relaxing this constraint, but
such fits are inconsistent with the data in the Fe\,K$\alpha$ regime, so
we apply this constraint throughout.  It is possible, of course, that the 
6.4\,keV line arises in photoionised gas, and not from reflection, and it is
also possible that there is a wide range of reflection ionisation 
(e.g. \citealt{ballantyne03}), but we shall start with simpler models to evaluate
the extent to which they describe the observations.  

Further constraints on the model are available from the soft band high resolution
\xmm {\sc rgs} and \chandra {\sc hetgs} grating data.  One particular feature
is that there is no evidence in any of the grating data for strong 
0.65\,keV\,\ion{O}{viii} emission, whereas the {\sc reflionx} models
lead us to expect significant line emission in the soft band to match the
observed 6.4\,keV\,Fe\,K$\alpha$ emission, if this arises from reflection.
So it must be that either the Fe\,K$\alpha$ line does not have a reflection
origin, 
or its ionisation is sufficiently low that no \ion{O}{viii} emission is expected,
or the reflection is affected by absorption in the soft band.  In
the models described here, we adopt the latter solution, although we should bear
in mind the former possibilities.  This interpretation of the Fe\,line leads to
the need to allow a further layer of absorption associated with the distant
reflection, {\em zone 4}, which may in fact be an atmosphere associated with
the reflection region.  This absorption zone hardens the reflection spectrum 
and thus for a given $2-10$\,keV flux increases the flux in the \suzaku {\sc pin}
band, and    
allowing the reflection to be ionised explains why the equivalent width of   
the Fe\,emission is lower than expected from neutral reflection
\citep{george91, zycki94, ross99}.
We should expect some emission from the absorbing gas in zone 4, with an equivalent
width of a few tens of eV \citep{leahy93}, more detailed models could include
this emission component also.  

We also allow a column of cold Galactic gas, as an additional free parameter
but constrained to have a minimum hydrogen column density of 
$4\times 10^{20}$\,cm$^{-2}$.  

\begin{table}
\caption
{Fit parameters and statistics for the joint fit of Model A to
eigenvector one and the offset components,
for each dataset.
First rows show photon index $\Gamma$, {\sc reflionx} log ionisation parameter
$\xi$, where $\xi$ is in units
of erg\,cm\,s$^{-1}$, 
and Galactic absorption 
hydrogen column density $N_H$ in units of $10^{22}$\,cm$^{-2}$.
The ionisation parameter and column for each of the four zones discussed
in the text follow, in the same units.  
Also shown is goodness-of-fit $\chi^2$ and
number of degrees of freedom, 
assuming a systematic fractional error of 0.03.
Brackets indicate parameters that were fixed.  
}
\label{table:pcafitsA}
\begin{center}
\begin{tabular}{lrrr}
\hline\hline
 & & \multicolumn{2}{c}{Model A} \\
\multicolumn{2}{c}{parameter} & \hspace*{5mm} \suzaku & \xmm \\
\hline
$\Gamma$     &            & 2.20 & 2.13 \\
\multicolumn{2}{l}{$\log\xi_{\rm REFLIONX}$}  & 2.08  & 2.08 \\
Galactic   & $N_H$        & 0.065 & 0.040 \\
\hline
\multicolumn{4}{l}{\chandra {\sc hetgs} \& \xmm {\sc rgs} zones}\\
{\em zone 1} & $\log\xi$   & 2.17  & 2.11 \\
             & $N_H$       & 1.44  & 0.46 \\
{\em zone 2} & $\log\xi$   & $-0.05$  & $-0.40$\\
             & $N_H$       & 0.01  & 0.02 \\
{\em zone 3} & $\log\xi$   & (3.85)  & (3.85) \\
             & $N_H$       & 3.00  & 10.1 \\
\hline
\multicolumn{4}{l}{additional offset component absorption zone}\\
{\em zone 4} & $\log\xi$   & 0.42  & 1.33 \\
             & $N_H$       & 4.88  & 8.33 \\
\hline
\multicolumn{2}{l}{$\chi^2$/dof} & 460/336 & 643/313\\
\hline
\end{tabular}
\end{center}
\end{table}

\begin{table}
\caption
{Fit parameters and statistics for the joint fit of Model B to
eigenvector one and the offset components,
for each dataset, table entries as in Table\,\ref{table:pcafitsA}.
}
\label{table:pcafitsB}
\begin{center}
\begin{tabular}{lrrr}
\hline\hline
 & & \multicolumn{2}{c}{Model B} \\
\multicolumn{2}{c}{parameter} & \hspace*{5mm} \suzaku & \xmm \\
\hline
$\Gamma$     &            & 2.23 & 2.21 \\
\multicolumn{2}{l}{$\log\xi_{\rm REFLIONX}$}  & 2.04  & 1.94 \\
Galactic   & $N_H$        & 0.051 & 0.045 \\
\hline
\multicolumn{4}{l}{\chandra {\sc hetgs} \& \xmm {\sc rgs} zones}\\
{\em zone 1} & $\log\xi$   & 2.36  & 2.14 \\
             & $N_H$       & 0.45  & 0.41 \\
{\em zone 2} & $\log\xi$   & 0.22  & $-0.42$ \\
             & $N_H$       & 0.07  & 0.04 \\
{\em zone 3} & $\log\xi$   & (3.85)  & (3.85) \\
             & $N_H$       & 2.60  & 2.90 \\
\hline
\multicolumn{4}{l}{additional offset component absorption zones}\\
{\em zone 4} & $\log\xi$   & 1.95  & 2.04 \\
             & $N_H$       & 34.0  & 28.1 \\
{\em zone 5} & $\log\xi$   & 1.35  & 1.89 \\
             & $N_H$       & 3.59  & 9.81 \\
\hline
\multicolumn{2}{l}{$\chi^2$/dof} & 261/332 & 257/309 \\
\hline
\end{tabular}
\end{center}
\end{table}

\begin{figure*}
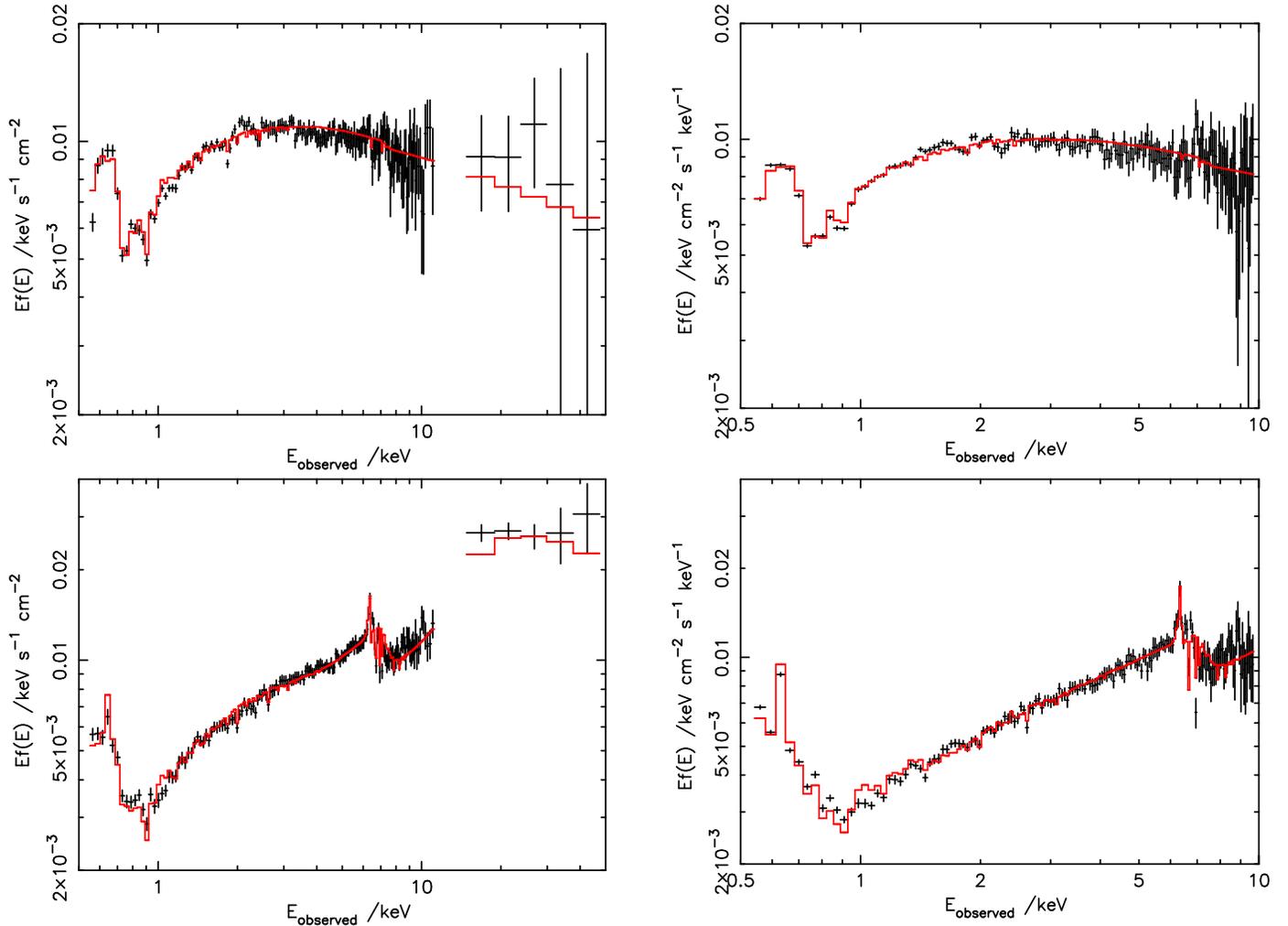

\begin{minipage}{\textwidth}{
\resizebox{0.475\textwidth}{!}{
\rotatebox{-90}{
\includegraphics{9590fig3a.ps}
}}
\hspace*{0.04\textwidth}
\resizebox{0.475\textwidth}{!}{
\rotatebox{-90}{
\includegraphics{9590fig3b.ps}
}}
}\end{minipage}

\begin{minipage}{\textwidth}{
\resizebox{0.475\textwidth}{!}{
\rotatebox{-90}{
\includegraphics{9590fig3c.ps}
}}
\hspace*{0.04\textwidth}
\resizebox{0.475\textwidth}{!}{
\rotatebox{-90}{
\includegraphics{9590fig3d.ps}
}}
}\end{minipage}

\caption
{The fit of model B, to the principal component spectra: (left) \suzaku 0.5-45\,keV,
(right) \xmm 0.4-10\,keV; (top) eigenvector one, (bottom) offset component.
The model is shown in units of Ef(E), points with
error bars show the `unfolded' component spectrum values.
}
\label{fig:pcafit}
\end{figure*}

This basic model based on the grating observations provides a 
qualitative fit to the principal components describing the variable spectrum of
MCG--6-30-15, and indicates that the absorbed, two-component source model is
a good basis for further investigation (Table\,\ref{table:pcafitsA}, model A.
In the results tabulated,
the {\sc reflionx} ionisation parameter reached its maximum allowed
value in the fit.
Zone 3 is assumed to absorb only
the offset component in fitting to the PCA components.
We do not at this stage quote statistical errors
on parameter values, deferring those until the model fits to the full
dataset).
Eigenvector one is well-fit in both datasets, 
although the fit of purely absorbed low-ionisation reflection to the
\xmm offset component is rather poor.

One problem that has already become clear with the general problem
of fitting reflection models to the \suzaku data is that if the hard
component is purely reflection, its amplitude at $\sim 20$\,keV requires a 
high reflected intensity, about three times that of the directly-viewed
power-law \citep{ballantyne03,miniutti07}.  
One solution to this problem is to suppose that
not all the hard offset component is purely reflection, but that it actually
is composed at least in part of an absorbed component.  Such an absorbed
component can appear in the offset component if its covering fraction is
variable (see the discussion in M07 and T07) - in this case the offset 
component effectively represents the appearance of the source in its lowest,
most covered, state.  
If we add an absorbed amount of the 
incident powerlaw onto both components (as might be required if this
arises as variable partial covering), with a new absorbing {\em zone 5},
it reduces the number of degrees of freedom by four 
and improves $\chi^2$ by 199 and 386 for the \suzaku and \xmm datasets,
respectively (Table\,\ref{table:pcafitsB}, model B): a substantial improvement
over the purely absorbed reflection model.
The amplitude of reflected emission is
reduced accordingly, thereby decreasing the requirement for an unusually
high reflected intensity.  If the reflector also has an uninterrupted view
of the full power-law component, this further decreases the ratio of reflected
to incident light.  This fit is shown 
in Fig.\,\ref{fig:pcafit} for both datasets.  However, 
eigenvector one becomes less easy to interpret in this
case, as it comprises contributions from both the direct continuum
and the variable-covering absorbed continuum.  For a robust determination of
the model parameters and for testing its goodness-of-fit it is therefore
essential that we fit the model directly to the data, as in the next section.

\section{Results - model fits to data}\label{datafitting}

\subsection{Fitting methodology and initial constraints}\label{models}
\begin{figure}
\begin{minipage}{\textwidth}{
\resizebox{0.475\textwidth}{!}{
\rotatebox{-90}{
\includegraphics{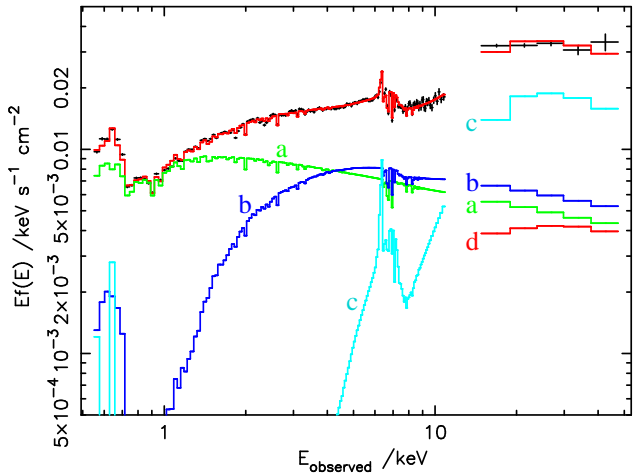}
}}
}\end{minipage}
\caption
{
Illustration of the spectral model.  The upper curve shows the model fitted to
the mean \suzaku spectrum, with {\sc xis} data below 11\,keV and {\sc pin}
data above 15\,keV.  Points with error bars show the unfolded data (see
section \ref{suzakufit} for details).
The three emission components are
shown as (a) primary directly-viewed power-law, absorbed by zones 1 \& 2;
(b) partially-covered power-law, absorbed by zones 1, 2, 3 \& 5;
(c) reflection, absorbed by zones 1, 2, 3 \& 4.  
In the fit to the PCA components, zone 3 is excluded from eigenvector one;
in the fit to the actual data, zone 3 is allowed to absorb all components.
Also shown is the 
expected contribution of the cosmic X-ray background to the \suzaku {\sc pin} band
(d) included in the model.
}
\label{model}
\end{figure}

Fitting to the PCA components should only be considered as yielding an indication
of the model components that may be required, for two reasons.  First, although
the offset component and eigenvector one alone provide a good description of the
variable X-ray spectrum at energies above 2\,keV (Tables\,\ref{table1}, \ref{table2}),
more complex source behaviour is implied at softer energies.  Second, there is no
unique interpretation of the offset component: this component essentially describes
the appearance of the source around its lowest possible flux state, and the 
offset component spectrum we deduce may either be pure reflection or pure absorption,
for a source with a variable absorption covering fraction, 
or some combination of the two, as indicated by the PCA.
Hence we need to test the model against the actual
data.  This is done in this section.
Model B is summarised by Fig.\,\ref{model} which shows the model
fitted to the mean \suzaku spectrum as described below and showing the three emission
components of the model 
(``direct'' power-law, with ionised absorption; ``partially-covered''
power-law, with higher opacity absorption from zone 5; 
and low-ionisation reflection with absorption from zone 4).
We also show the component of cosmic X-ray background emission included in the
fit to {\sc hxd pin} data.

\begin{figure}
\begin{minipage}{\textwidth}{
\resizebox{0.475\textwidth}{!}{
\rotatebox{-90}{
\includegraphics{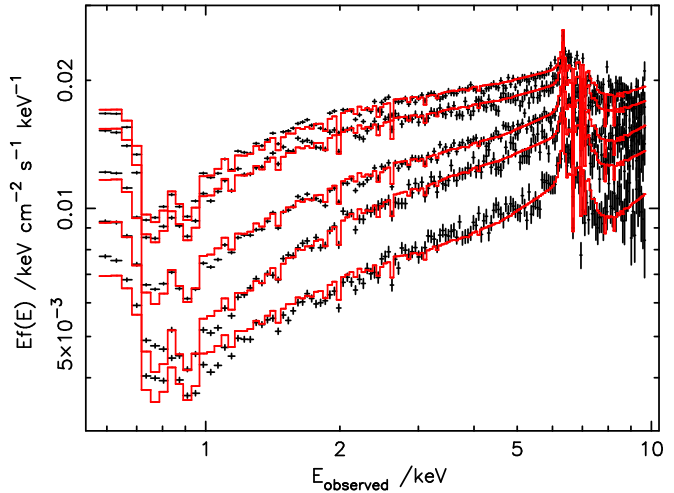}
}}
}\end{minipage}
\caption
{Simultaneous fits to the \xmm data split into 5 flux states, 0.55-9.7\,keV.
The model is shown in units of Ef(E), points with
error bars show the `unfolded' data.
}
\label{xmmfluxstates}
\end{figure}

The full dataset that we investigate here comprises observations taken at a 
number of epochs with a variety of instruments.
In fitting to the data we require the model to fit simultaneously any data
that were obtained simultaneously (e.g. {\sc rgs} and {\sc pn} data for
\xmm or {\sc pin} and {\sc xis} data for \suzaku).  We do allow variations
in model parameter values between datasets taken at different epochs, although
the model components are not changed.  

Some model components are better constrained by some datasets than others.  A
particular case is the high-ionisation outflowing 
zone 3, which is most strongly constrained
by the high-resolution grating data, and in particular by the \chandra {\sc heg}
data around 6.7\,keV \citep{young05}.  In order to ensure that the models fitted
to CCD-resolution data are consistent with the \chandra grating dataset, we first
estimate parameters for zone 3 by fitting a simple model to the \chandra data 
that reproduces the 
outflowing 6.7\,keV and 6.97\,keV line equivalent widths and redshift.  Those
model parameters are then fixed in fits to the other datasets.  Once the 
other datasets have also been fitted, the full model is then also fitted to
the \chandra grating data as a final check on the model.

Because these
lines are easily saturated, the equivalent widths are strongly dependent on the 
assumed turbulent velocity dispersion in the absorbing gas.  
A low turbulent velocity would require
a high column, close to Compton-thick, in order to obtain the equivalent widths
observed.  A higher turbulent velocity dispersion would lead to line widths
inconsistent with those observed (these statements were tested using 
{\sc xstar} models created with velocity dispersion
$\sigma = 10, 300, 500, 1000$\,km\,s$^{-1}$).  
We use a model with $\sigma = 500$\,km\,s$^{-1}$.  
Thus the column and ionisation fit parameters we obtain would need to
be modified if the true $\sigma$ differs from 500\,km\,s$^{-1}$, but
a comparable goodness of fit can be obtained.  We find higher velocity
dispersions need lower columns at fixed ionisation parameter to reproduce
the observed line equivalent widths, but that
higher values of ionisation parameter are needed to reproduce the
6.7, 6.97\,keV line ratios.
The initial absorber parameter values found from the \chandra data were
$N_{\rm H}=2.1\times10^{22}$\,cm$^{-2}$, $\log\xi=3.85$
and outflow velocity 1800\,km\,s$^{-1}$ with respect to the systemic redshift.
The same value of $\xi$ and the outflow velocity were adopted when fitting to
the PCA components.  

In fitting to the PCA, we observed that the zone 3 absorption lines 
appeared only on the offset component, which implies that either
(i) the absorber is only in front of the region responsible for the offset
component, or (ii) the equivalent widths of the zone 3 lines decrease with
increasing flux, perhaps because of increasing ionisation.  The same effect
is seen in Mrk\,766 (T07).  In fitting to the data we initially adopt the
same zone 3 column and ionisation for all flux states of the source, but later
we shall investigate fits in which the ionisation of zone 3 is allowed to
vary between different flux states.

\subsection{Fits to \xmm multiple flux states}\label{secfluxstates}

The model is now compared with the variable X-ray spectrum more directly
than in the PCA, 
by dividing the data into a number of ``flux states'' each with a different
spectrum, and fitting jointly to those spectra.  We start with the \xmm data.

\begin{figure*}
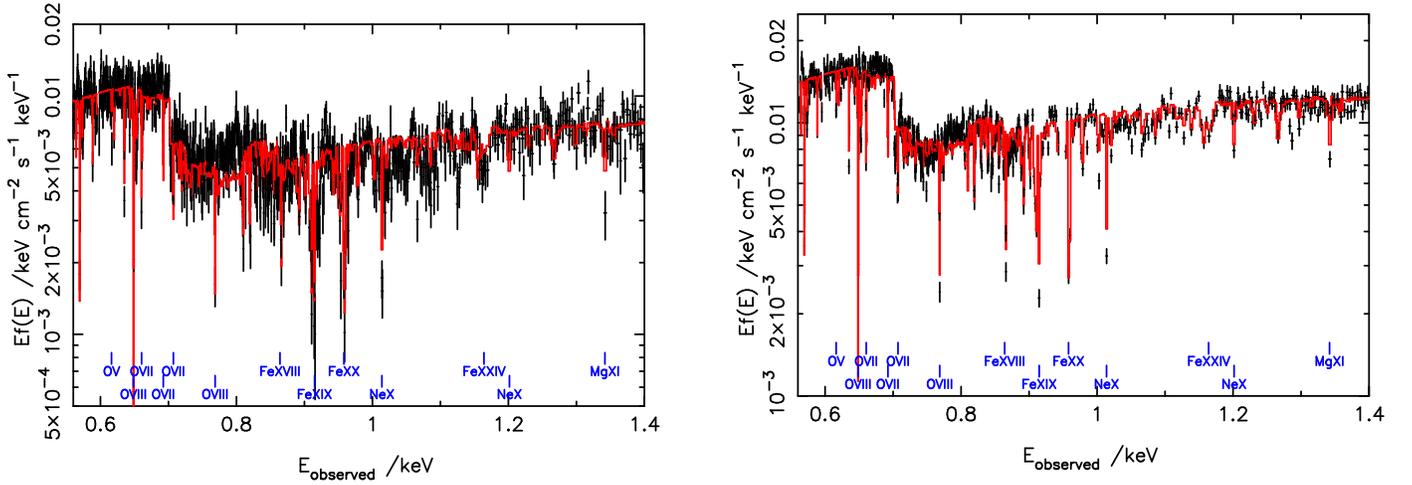

\begin{minipage}{\textwidth}{
\begin{minipage}{0.475\textwidth}{
\resizebox{\textwidth}{!}{
\rotatebox{-90}{
\includegraphics{9590fig6a.ps}
}}}\end{minipage}
\hspace*{0.04\textwidth}
\vspace*{0.05\textwidth}
\begin{minipage}{0.475\textwidth}{
\resizebox{\textwidth}{!}{
\rotatebox{-90}{
\includegraphics{9590fig6b.ps}
}}}\end{minipage}
}\end{minipage}
\caption
{
Fit to (left) the 2000 and (right) the 2001 
\xmm {\sc rgs} spectra of MCG--6-30-15:
The model is shown in units of Ef(E), points with
error bars show the `unfolded' data.
}
\label{figrgs}
\end{figure*}

\begin{table}
\caption{Model parameter values for the \xmm {\sc pn + rgs}, 
\suzaku {\sc xis + pin} and \chandra {\sc meg + heg} 
fits,
where $\Gamma$ is power-law photon index, $\xi$ is ionisation parameter in units
of erg\,cm\,s$^{-1}$ and N$_{\rm H}$ is absorber hydrogen column density assuming
solar abundances, in units of $10^{22}$\,erg\,cm\,s$^{-1}$. 
Quoted uncertainties are 68\,percent confidence intervals.
Brackets indicate values that were not free parameters for that particular dataset.
The final rows give goodness-of-fit:
the data for each mission were jointly fit, but the contributions to $\chi^2$
are quoted separately for each of the subsets of \xmm data: 
$^a$ {\sc pn}, $^b$ 2001 {\sc rgs}, $^c$ 2000 {\sc rgs} (see text).
}
\label{parvals}
\begin{tabular}{lr@{ $\pm$ }lr@{ $\pm$ }lr@{ $\pm$ }l}
\hline\hline
parameter & \multicolumn{2}{c}{\xmm} & \multicolumn{2}{c}{\suzaku} & \multicolumn{2}{c}{\chandra} \\
\hline
$\Gamma$             & 2.284 & 0.013    & 2.265 & 0.017 & \multicolumn{2}{c}{(2.284)} \\
$\log\xi_{\rm REFLIONX}$ & 2.04 & 0.01 & 1.97  & 0.03  & \multicolumn{2}{c}{(2.04)} \\
N$_{\rm H}$(Gal)     & 0.040 & 0.004   & 0.052 & 0.003 & 0.057 & 0.002 \\
$\tau_{\rm 0.7\,keV\,edge}$ & 0.42 & 0.02   & 0.36  & 0.03  & 0.45 & 0.03 \\

N$_{\rm H}$(1)       & 0.26 & 0.02    & 0.45  & 0.05  & 0.11 & 0.08 \\
$\log\xi$(1)         & 2.64 & 0.02    & 2.78  & 0.07  & 2.33 & 0.05 \\
N$_{\rm H}$(2)       & 0.016 & 0.001   & 0.03  & 0.001 & 0.022 & 0.001 \\
$\log\xi$(2)         & 0.25 & 0.08   & $-0.11$ & 0.1   & $-0.04$ & 0.16 \\

N$_{\rm H}$(3)       & \multicolumn{2}{c}{(2.10)} & \multicolumn{2}{c}{(2.10)} & \multicolumn{2}{c}{(2.10)} \\
$\log\xi$(3)         & \multicolumn{2}{c}{(3.85)} & \multicolumn{2}{c}{(3.85)} & \multicolumn{2}{c}{(3.85)} \\

N$_{\rm H}$(4)      & 34.8 & 0.05    & 54.9  & 0.05  & \multicolumn{2}{c}{(34.8)} \\
$\log\xi$(4)        & 1.83 & 0.01    & 1.94  & 0.01  & \multicolumn{2}{c}{(1.83)} \\

N$_{\rm H}$(5)      & 5.40 & 0.08    & 4.32  & 0.11  & \multicolumn{2}{c}{(5.40)} \\
$\log\xi$(5)        & 1.75 & 0.02    & 1.39  & 0.05  & \multicolumn{2}{c}{(1.75)} \\
\hline
flux-states & \multicolumn{2}{c}{$799/783^{a}$} & \multicolumn{2}{c}{715/841} & \multicolumn{2}{c}{} \\
$\chi^2$/dof & \multicolumn{2}{c}{} & \multicolumn{2}{c}{} & \multicolumn{2}{c}{} \\
\hline
mean spectra & \multicolumn{2}{c}{$102/159^{a}$} & \multicolumn{2}{c}{115/171} & \multicolumn{2}{c}{3406/3341} \\
$\chi^2$/dof             & \multicolumn{2}{c}{$1965/873^{b}$} & \multicolumn{2}{c}{}&\multicolumn{2}{c}{} \\
             & \multicolumn{2}{c}{$1820/1074^{c}$} & \multicolumn{2}{c}{}&\multicolumn{2}{c}{} \\
\hline 
\end{tabular}
\end{table}

The \xmm {\sc pn} data were each divided
into 5 flux states, separated by equal logarithmic intervals of flux defined
in the 1-2\,keV band, spanning the range of flux covered by the data.
These states were fit simultaneously, with a single
value of parameters such as power-law index and absorber 
column density and ionisation
parameters, with
absorber properties initially chosen to match the fits to the PCA.  
The optical depth of the dust edge was also allowed to be a single free
parameter, the same for all flux states.
For each flux state,
the direct, absorbed power-law and reflected component amplitudes were allowed to vary
independently. Hence there were 15 free parameters, of which three were allowed 
to vary between flux states.  We fix abundances to solar values throughout.

\begin{figure}
\resizebox{0.475\textwidth}{!}{
\rotatebox{-90}{
\includegraphics{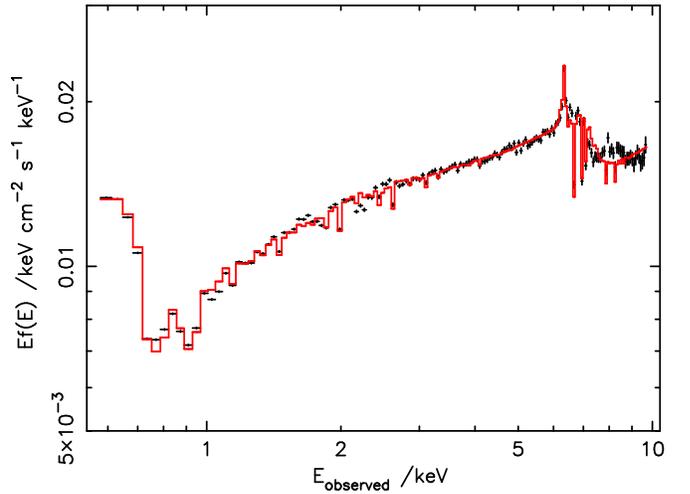}
}}
\caption
{
Fit to the mean \xmm spectrum of MCG--6-30-15.
The model is shown in units of Ef(E), points with
error bars show the `unfolded' data.
}
\label{figxmmmean}
\end{figure}

\begin{figure}
\begin{minipage}{\textwidth}{
\resizebox{0.475\textwidth}{!}{
\rotatebox{-90}{
\includegraphics{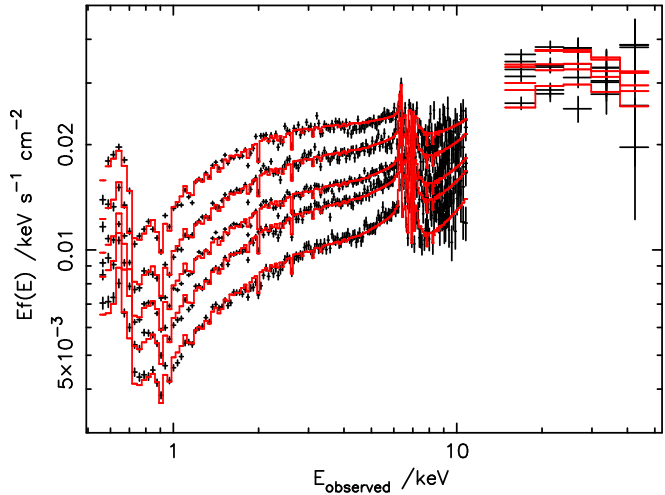}
}}
}\end{minipage}
\caption
{Simultaneous fits to the \suzaku data split into 5 flux states, 0.5-45\,keV.
The model is shown in units of Ef(E), points with
error bars show the `unfolded' data.
}
\label{suzakufluxstates}
\end{figure}

As discussed above, it is important to further constrain the soft-band emission
from the models, to be consistent with the high-resolution grating data. The
\xmm observations comprised simultaneous {\sc rgs} and {\sc pn} observations,
and hence the models were fit simultaneously to the {\sc pn} flux state and
the {\sc rgs} data.  Because of low signal-to-noise, the {\sc rgs} data were
not themselves divided into flux states, and the model was required to fit the
mean {\sc rgs} spectrum.  Because of its limited energy range, the amplitudes of
the various components are not well constrained by the {\sc rgs} data, 
so as well as the individual {\sc pn} flux state data, the {\sc pn} mean 
spectrum was also included in the fit and the {\sc rgs} parameters were tied to
the parameters for that component.  A nominal cross-calibration constant factor was allowed
for each of {\sc rgs 1} and {\sc rgs 2}
between the {\sc rgs} and {\sc pn} fits, although the value of this parameter
was unity to within one percent.
The 2001 {\sc rgs} data are of significantly higher
signal-to-noise than the 2000 data, so only the 2001 data was used to constrain
the model parameters using the above procedure.
Having obtained the absorber and emitter parameters, 
we present also the results of jointly fitting the same model to the 2000 data,
with the same absorber parameters and only differing normalisations of the emission
components.

This joint analysis yields an overall goodness-of-fit $\chi^2 = 2773$ for
1689 degrees of freedom.  The excess in $\chi^2$ arises entirely in the
{\sc rgs} portion of the data, although the model still describes
the high-resolution spectrum with an accuracy about 5\,percent and
correctly reproduces the chief lines and edges seen in the {\sc rgs} data. 
The model fit to the {\sc rgs} portion
is discussed more below and in section\,\ref{abundances}.

Fig.\,\ref{xmmfluxstates} shows the fit to the {\sc pn} data portion.
The contribution to $\chi^2$ from the {\sc pn} data alone is
$\chi^2 = 799$ for 783 degrees of freedom (Table\,\ref{parvals})
(counting all free parameters when quoting the number of degrees of freedom).
Parameter uncertainties quoted in the table are 68\,percent confidence intervals.
We note that this is not the best fit that
can be achieved if the {\sc rgs} data are ignored:  but the fit to the
{\sc pn} data that is constrained by both
{\sc pn} and {\sc rgs} data is nonetheless a good fit.
The largest discrepancy between the {\sc pn} data 
and the model occurs in the soft band in the
lowest flux state, and likely indicates the effect of time variable
absorption, which we have not taken into account in this analysis.

\begin{figure}
\hspace*{-0.04\textwidth}
\begin{minipage}{\textwidth}{
\vspace*{-0.03\textwidth}
\resizebox{0.53\textwidth}{!}{
\rotatebox{-90}{
\includegraphics{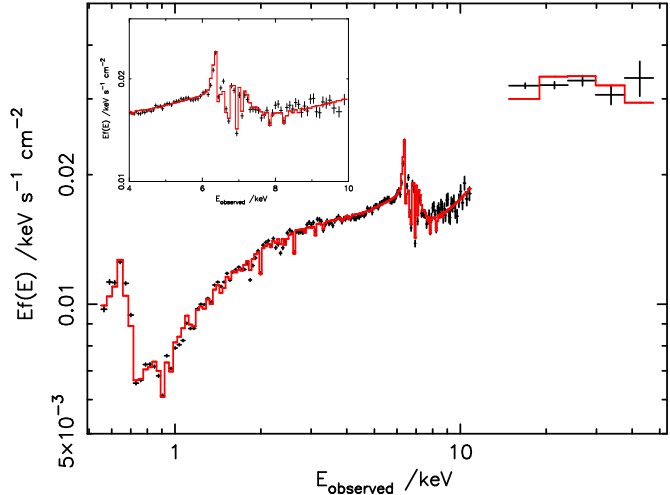}
}}
}\end{minipage}
\caption
{
Fit to the mean \suzaku spectrum of MCG--6-30-15:
with inset showing the zoom-in to the
4-10\,keV region.
The model is shown in units of Ef(E), points with
error bars show the `unfolded' data.
}
\label{figmean}
\end{figure}

Fig.\,\ref{figrgs} shows the 2001 {\sc rgs} portion of the model fit, and
also shows the fit of the same model to the 2000 {\sc rgs} data as mentioned above.
The {\sc rgs 1} and {\sc rgs 2} data are overplotted on the figure rather than being
averaged together in regions of overlap.  The models fit less
well than for the {\sc pn}, with the {\sc rgs} contribution to $\chi^2$ being
$\chi^2 = 1965$ for 873 degrees of freedom for the 2001 data in the energy range
shown, $0.56-1.4$\,keV (to be conservative, again, all jointly-fit parameters are counted
as being free for the calculation of the number of degrees of freedom).
The 2000 data are of lower signal-to-noise and as a consequence have a better
goodness-of-fit, $\chi^2 = 1820$ for 1074 degrees of freedom in the same
energy range.
Note that, as above, the quoted fit is not the best-fit of the model to the {\sc rgs}
data alone, but rather is the contribution to $\chi^2$ of the {\sc rgs} data
to the joint {\sc pn} and {\sc rgs} best fit, and therefore is degraded by
any cross-calibration uncertainty between the two instruments.
Given also the complexity of the {\sc rgs} spectra 
and the relative simplicity of the models, including the assumption of solar abundances
(section\,\ref{abundances}),
these relatively poor {\sc rgs} fits are not too surprising.
The largest discrepancies between model and data in the energy range considered are
at energies just above the 0.7\,keV edge, 
and the simple, single-edge model that we have adopted
is likely too simplistic.  We do not have sufficient information to 
attempt more sophisticated models of this region.
However, the reproduction of the overall continuum shape, the lack of strong
emission lines and the good correspondence with the 
majority of the absorption features indicates that
the small number of zones that have been included do account for the majority
of the features.  
Many of the absorption lines identified by \citet{turner04} are reproduced
by the model, adopting those authors' line identifications we find a good
match for 
0.65\,keV\,\ion{O}{viii}\,Ly$\alpha$, 
0.77\,keV\,\ion{O}{viii}\,Ly$\beta$, 
0.87\,keV\,\ion{Fe}{xviii},
0.92\,keV\,\ion{Fe}{xix},
0.96\,keV\,\ion{Fe}{xx},
1.02\,keV\,\ion{Ne}{x}\,Ly$\alpha$ and
1.35\,keV\,\ion{Mg}{xi}.
These lines chiefly originate in zone 1, and some from zone 3, 
with zone 2 providing most of the 
lines and edges of \ion{O}{v}-\ion{O}{vii}.

We also fit the same additive model to the mean \xmm spectrum
where model parameters were fixed at the values found above,
only allowing the normalisations of the three emission components to float.
This results in a somewhat better-than-expected fit (assuming systematic fractional
error 0.03), $\chi^2=102$ with 159 degrees of freedom 
(Table\,\ref{parvals}, Fig.\,\ref{figxmmmean}),
although of course the same data were used to generate the model with more
free parameters when analysing the multiple flux state data above.

\subsection{Fits to \suzaku multiple flux states and mean spectrum}\label{suzakufit}
We carry out a similar procedure for fitting to the \suzaku data, except now
there is no simultaneous high resolution data, but there is simultaneous
{\sc xis} and {\sc pin} data.
The flux-state XIS data were noisy above 10.5\,keV, so the XIS data were truncated 
at that energy.
When fitting to the \suzaku {\sc pin}
data, the data were not corrected for the contribution of the cosmic X-ray
background (CXB), but instead the CXB model adopted by \citet{miniutti07} was
included in the fit, with a normalisation allowed to float by $\pm 5$\,percent
to allow for the uncertainty in the model and in the \suzaku absolute calibration.

The results are displayed in 
Fig.\,\ref{suzakufluxstates}.  Fitting to all five flux states with this model
yields $\chi^2 = 715$ for 841 degrees of freedom, best-fitting parameter values
are given in Table\,\ref{parvals}.

We also fit the same additive model to the mean \suzaku spectrum
where again the
model parameters were fixed at the values found in the flux states analysis,
only allowing the normalisations of the three emission components to float.
This results in goodness-of-fit 
$\chi^2=115$ for 171 degrees of freedom
(Table\,\ref{parvals}, Fig.\,\ref{figmean}).  
An even lower value of $\chi^2$ may be obtained by allowing 
other parameters to float, but as the model would then no longer fit the variable
flux state data we do not allow this freedom.

\begin{figure}
\resizebox{0.475\textwidth}{!}{
\rotatebox{-90}{
  \includegraphics{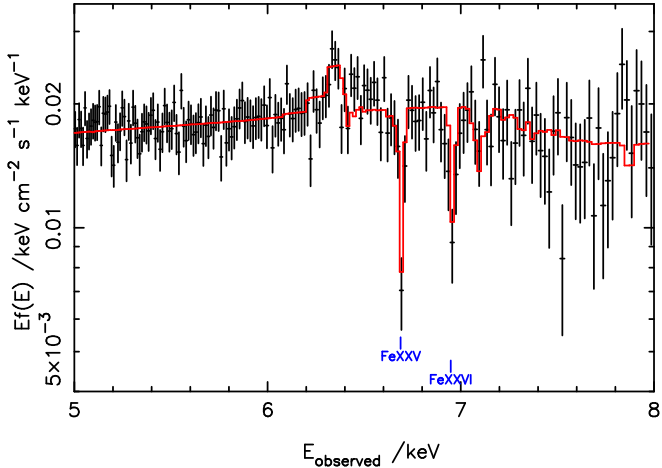}
  }}
\caption{Model fit to \chandra {\sc heg} spectrum in the region $5-8$\,keV 
(solid line) with unfolded data shown as points with errors.
}
\label{fig:heg}
\end{figure}

\begin{figure*}
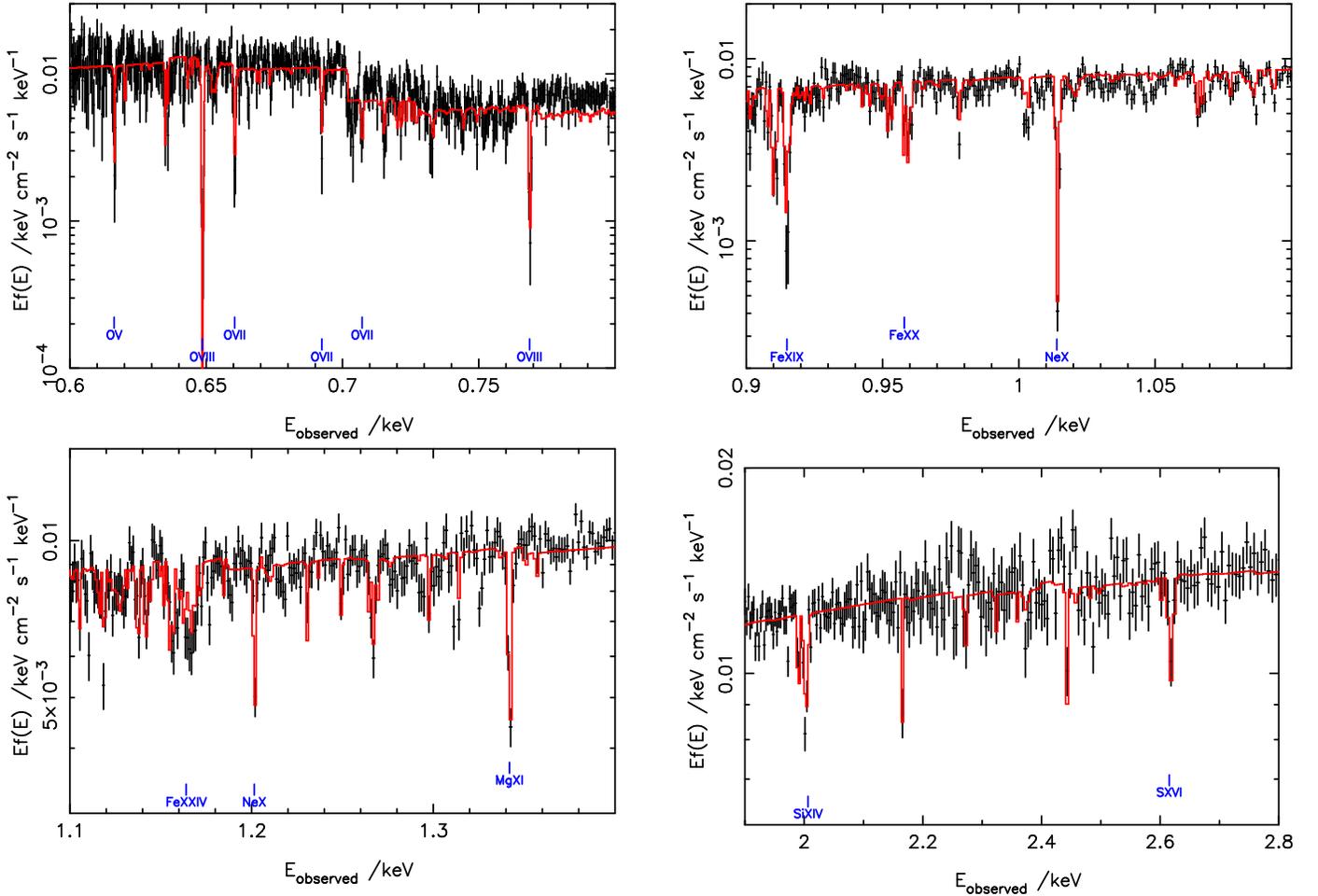

\begin{minipage}{\textwidth}{

\resizebox{0.475\textwidth}{!}{
\rotatebox{-90}{
\includegraphics{9590fig11a.ps}
}}
\hspace*{0.04\textwidth}
\resizebox{0.475\textwidth}{!}{
\rotatebox{-90}{
\includegraphics{9590fig11b.ps}
}}

\resizebox{0.475\textwidth}{!}{
\rotatebox{-90}{
\includegraphics{9590fig11c.ps}
}}
\hspace*{0.04\textwidth}
\resizebox{0.475\textwidth}{!}{
\rotatebox{-90}{
\includegraphics{9590fig11d.ps}
}}

}\end{minipage}
\caption{
Model fit to the \chandra {\sc meg} data, showing selected energy ranges,
with the model shown as the solid line, unfolded data as points with error bars.
}
\label{fig:meg}
\end{figure*}

\subsection{Fits to \chandra {\sc hetgs} data}
The available \chandra {\sc hetgs} data has previously been analysed by 
\citet{lee01} and \citet{young05}
who have demonstrated its diagnostic power in detecting the presence 
of absorbing layers.  
The data were taken at a different epoch from the \xmm and
\suzaku data, with a different instrument, and so we expect to need to vary some
of the parameters in order to obtain a good fit.  
The mean {\sc heg} spectrum over the range $3-8$\,keV and 
the mean {\sc meg} spectrum over the range $0.55-5$\,keV
were jointly fit with the same model as the other datasets, with free
parameters as before, except that the power-law slope, {\sc reflionx}
ionisation parameter and some absorption zones
were fixed at the \xmm values (see Table\,\ref{parvals}; the mean \chandra
spectrum on its own does not sufficiently constrain these parameters).
The resulting parameter values
are shown in Table\,\ref{parvals}, which resulted in a goodness-of-fit
$\chi^2=3406$ for 3341 degrees of freedom for the joint fit.

Viewing the fit in more detail, 
we first consider the {\sc heg} data in the region around 7\,keV.  The pair of
absorption lines at this energy have already been discussed in this paper and
by \citet{young05, brenneman06} and \citet{miniutti07}.  Fig.\,\ref{fig:heg}
shows the model fitted to the data, and it can be seen both that the absorption
lines are reproduced by this model and that no other strong lines are expected,
and it is therefore consistent with the constraint discussed by
\citet{young05} that there should not be significant amounts of absorption from
intermediate ionisation states of Fe, which would produce absorption at
$\sim 6.5$\,keV.  This constraint is avoided because the red wing is largely
produced by zone 5, which has ionisation sufficiently low that the Fe\,L shell
is filled, at $\log\xi \la 2$ \citep{kallman04}, and it is the partial covering
of this zone that allows the continuum shape to be correctly reproduced as well
as explaining the variability properties (discussed further in 
section\,\ref{partialcovering}).

Further confirmation of the high-ionisation layer comes from the \chandra
{\sc meg} data, in which numerous absorption features arising in this layer are
visible, including the features at 2.0 and 2.62\,keV discussed by \citet{young05}
(Fig.\,\ref{fig:meg}).  

As in the {\sc rgs} data, the {\sc meg} data reveal
numerous absorption lines in the soft band, many of which were previously
identified by \citet{lee01} and \citet{turner04}, including
0.615\,keV\,\ion{O}{v},
0.635\,keV\,\ion{O}{vi},
0.66\,keV and 0.69\,keV\,\ion{O}{vii},
0.65\,keV\,\ion{O}{viii}\,Ly$\alpha$, 
0.77\,keV\,\ion{O}{viii}\,Ly$\beta$, 
0.92\,keV\,\ion{Fe}{xix},
1.02\,keV\,\ion{Ne}{x}\,Ly$\alpha$,
1.35\,keV\,\ion{Mg}{xi} and 
2.0\,keV\,\ion{Si}{xiv}\,Ly$\alpha$
(we again adopt the line identifications of \citet{lee01} and \citet{turner04}).
Again, the higher ionisation lines chiefly originate in zones 1 and 3, the
lower ionisation lines in zone 2.

\subsection{The effect of non-solar abundances}\label{abundances}
In the above models, solar abundances were assumed throughout.  However, there is
evidence in MCG--6-30-15 for departures from those values: \citet{turner04}
point out that the observed equivalent widths of absorption lines in their
model are too low by a factor $\sim 1.25$, an effect that is also clearly seen
in the fits shown here (e.g. Fig.\,\ref{figrgs}).  The effect could be explained
either by altering the assumed velocity dispersion, or by altering the assuming abundance.Clearly, allowing either of these parameters to float would result in an overall
goodness of fit better than presented here.  Because of the degeneracies in such
fits with other parameters, not only velocity dispersion but also column density
and ionisation parameter, the current data is insufficient to
unambiguously determine absolute values of element abundances.  We can, however,
estimate the size of the effect of changing the assumed abundance to an alternate
value.  \citet{turner04} suggest O may need to be enhanced by a substantial
factor, perhaps as high as $3-4$, to
explain the observed equivalent widths.  We have therefore replaced the absorber
model used for zone\,1 with a new
{\sc xstar} model with $\alpha$-element abundance enhanced by a factor two, leaving
other abundances fixed at solar.  Consistent with the inference of \citet{turner04},
we find an improved overall fit to the \xmm combined {\sc pn}/2001\,{\sc rgs} data.
The overall $\chi^2$ for the combined fit shows a modest 
improvement by $\Delta\chi^2 = 39$, from a value 2773 to 2734 
(with 1689 degrees of freedom in both cases).
As expected, the equivalent widths of the O, Ne and Mg 
lines match better to the data, but the Fe lines remain with model equivalent widths
that are too small.  It seems likely therefore that there is an overall enhancement of
all elements, not only the $\alpha$ elements.  

The chief aim of this paper is to investigate the extent to which an absorption model
can describe the full X-ray spectrum of MCG--6-30-15.  In order not to confuse 
the general properties of the absorption model with specific details about the
likely element abundances, we do not here investigate further the more detailed effect
of non-solar abundances.

\subsection{Variation in absorber ionisation}\label{ionvar}
\begin{figure}
\begin{minipage}{\textwidth}{
\resizebox{0.475\textwidth}{!}{
\rotatebox{-90}{
\includegraphics{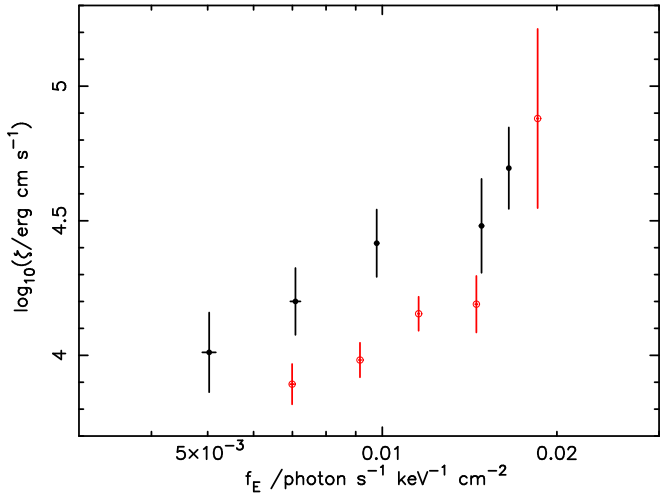}
}}
}\end{minipage}
\caption
{
The fitted variation in $\xi$ as a function of normalisation of the primary
power-law, for the \xmm (solid points) and \suzaku (open points) flux-states.
}
\label{xivar}
\end{figure}

We can also test the absorber models for possible variation in ionisation
parameter with flux.  There is evidence for this in the high ionisation
outflow, zone 3, in that the equivalent width of the absorption lines decrease
with increasing source brightness (equivalently, they only appear on the PCA
offset component, and not on eigenvector one).  This might arise if the zone 3
absorber is localised to the region responsible for the offset component,
but the alternative explanation investigated here is that the variation
in equivalent width is an effect of varying ionisation.  
Fig.\,\ref{xivar} shows the variation in $\xi$ for
each of the five \xmm and \suzaku flux states when this is allowed to vary
between each state.  The goodness-of-fit improves 
by $\Delta\chi^2 = 15$ to $\chi^2=822$ for 838
degrees of freedom: in itself this is a weak return for introducing a further
four free parameters, but it does reveal the expected relationship between
ionisation and flux. Within each dataset there is
a clear trend for $\xi$ to be proportional to the continuum amplitude, although
the \suzaku points appear offset to lower ionisation, implying a change in
either absorber density or ionising continuum spectrum between 2001 and 2006.

\section{Discussion}

\subsection{Model complexity and uniqueness}
The model presented above provides a good description of the variable X-ray
spectrum of MCG--6-30-15 throughout all its flux states, and correctly
reproduces the hard low-state and softer high-state spectra.  It also
reproduces the high energy $15-40$\,keV flux and its relative lack of variability,
and removes the requirement for an unexpectedly high reflection albedo.
The model is complex, requiring five intrinsic absorption zones, plus a dust
edge, but we know already from the \chandra and \xmm grating data that multiple
zones covering a wide range of ionisation, with at least two kinematically
distinct regions, are required by the data.  We should not be surprised by the
complexity of the model, three of the zones have been detected by previous
authors, but of course the next step in proving or disproving the
specific model presented here would be to search for high resolution signatures
of the additional heavy absorbing zones that are required.  
For the ionisation predicted from the model in this paper, there may
be a weak signature of the Fe\,K$\beta$ UTA arising from
ionisation states less ionised that \ion{Fe}{xvii}.  
This would be too weak to detect in the
present data but may be a detectable signature in data from
future missions with calorimeter detectors.

We might compare the model
complexity with other studies: \citep{brenneman06}, for example, fit the spectrum
expected from close to a Kerr black hole to the \xmm data for MCG$-6$-30-15,
including some of the ionised absorbing zones and the dust edge, over the range
$0.6-10$\,keV, but only fitting the mean spectrum.  That model does not include
all the ionised zones seen with \chandra, but still has a total of 18 parameters, 
of which eight are associated with the relativistically
blurred line component, and their model 4 results in a goodness-of-fit $\chi^2=1742$ for
1375 degrees of freedom.
It seems that whatever physical premise
is taken for the origin of the red wing, models of some complexity are needed.

Conversely, 
although the absorption model presented here has been successful, we cannot claim
that it is unique.  In this paper we have concentrated on a model in
which absorption plays a dominant role, and have developed the model constituents
in a systematic manner based on fits to the variable-spectrum low resolution data
and the high-resolution grating data.  This exercise at least shows that it
is possible to explain the X-ray spectrum of MCG--6-30-15 by such a model.
Many previous studies have concentrated solely on modelling the hard, low-variability
``offset'' component as being relativistically-blurred reflected emission from the 
inner accretion disc 
(\citealt{wilms}, \citealt{vaughanfabian04}, \citealt{reynolds04}, \citealt{miniutti07},
\citealt{brenneman06}), but 
those studies have only tested that model against the mean spectra, albeit of
data obtained when the source was in differing flux states, and often over a 
restricted range in energy (e.g., $3-45$\,keV in the study of \citealt{miniutti07}).
The model presented in this paper does not have any such blurred component,
but this does not prove that such a component does not exist.  After all,
an arbitrarily low amplitude of such a component could always be added into the
model described in this paper.  Nonetheless, the model-fitting presented here
does show that the offset component need not be considered as being {\em dominated}
by relativistically-blurred reflection.  One important consequence is that
it is not possible to use the spectral shape of the red wing 
in the current generation of data to deduce parameters such as black hole spin.  
Given the extreme
spin parameters that are deduced for AGN such as MCG$-6$-30-15
(in this case $a=0.989_{-.002}^{+.009}$, \citealt{brenneman06})
this has important consequences for models of black hole evolution, as
``chaotic'' mergers produce lower expected spin parameter values, and such 
extreme high values are expected only to be achieved by long-lived steady accretion
\citep{berti08,king08}.

\subsection{The role of absorber partial covering}\label{partialcovering}
The absorbed power-law component that forms part of the model presented here
is key to both understanding the soft-band continuum shape and the high flux
observed in the {\sc pin} data at high energy.  

Many previous authors have suggested that absorption partially covers an
X-ray source and that variations in covering fraction of an absorbing zone
may be linked to flux and spectral variations in both AGN and
Galactic black hole systems 
\citep[e.g.][{\em inter alia}]{holt80, reichert85, boller02, boller03, 
immler03, tanaka04, gallo04a, gallo04b, pounds04, turnerea05, grupe07}.
\citet{vaughanfabian04} previously tested a partial
covering model for the 2001 \xmm spectrum of MCG$-6$-30-15 and concluded that, 
although such a model provided an acceptable fit, a relativistically-blurred
model was superior.  The model tested in that case was solely a neutral
absorber however, a key difference with the models tested here.
\citet{mckernan98} also suggested specifically for
MCG--6-30-15 that a sharp dip in flux was caused by occultation by an
absorber, an idea revived most recently for NGC\,3516 by \citet{turner08}.
Mrk\,766 shows extremely similar X-ray spectral variability to MCG--6-30-15
and M07 and T07 have suggested that variable partial covering may play an
important or perhaps even dominant role in the X-ray spectral variability
of this source.  

\begin{figure}
\begin{minipage}{\textwidth}{
\resizebox{0.475\textwidth}{!}{
\rotatebox{-90}{
\includegraphics{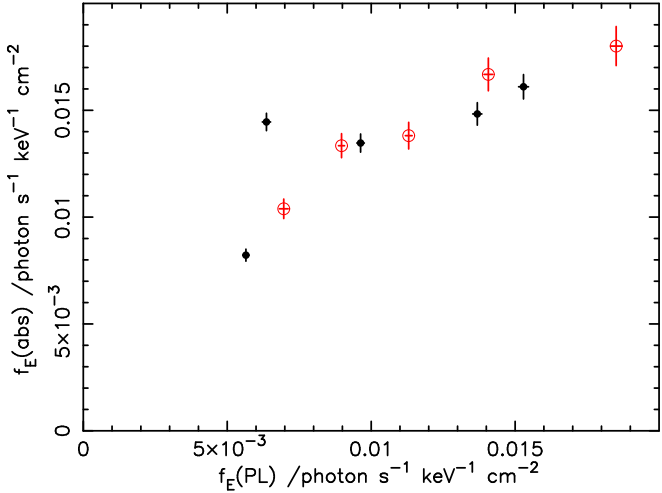}
}}
}\end{minipage}
\caption{The variation in amplitude of the absorbed component (y-axis) as a 
function of the amplitude of the direct power-law component (x-axis), for the
fits to the \suzaku (open symbols) and \xmm (solid symbols) data.}
\label{absorber-variations}
\end{figure}

To investigate this further we show in Fig.\,\ref{absorber-variations} the
amplitude in the absorbed component compared with the amplitude in the direct
component, as measured from the normalisation of the power-law in each
case.  It may be seen that, in the model presented here, the absorbed component
does vary coherently with the direct continuum at high flux states, but 
not with a dependence that passes through the origin.  
We can see from this diagram why the PCA results in an offset component that
contains a signficant amount of absorbed continuum: a linear extrapolation
through the points in Fig.\,\ref{absorber-variations} would hit the 
y-axis at a positive value of the absorbed component flux.  
At the lowest observed
fluxes, the correlation seems to break down, with a larger scatter in the
two components.
The simplest
interpretation of the observed trends is that in the highest flux states the
source has a covering fraction around 50\,percent, but that
at lower flux states the covering fraction is more variable and may increase
towards 100\,percent.  If this interpretation is correct it implies that the
covering fraction is a function of the flux state of the source and perhaps
indicates a dependence of either source size or absorber extent on flux state.
The trends seen here and the interpretation is however model
dependent, and at this stage we can do no more than suggest this interpretation.

\subsection{Hard-band reflection}
A problem that any model must address is the relatively high flux observed 
above 20\,keV \citep{ballantyne03,miniutti07}.  In the model presented here,
some fraction of the hard-band flux is still provided by distant reflection.
If that reflection has a view of the entire unabsorbed source output, then 
the reflected intensity relative to that expected from a disc subtending
2$\pi$\,sr (the ``R'' parameter) has a value around 1.7 (estimated as in
\citealt{miniutti07} from a comparison of component fluxes in the hard band):
still greater than unity but substantially smaller than $R \ga 3$ as
required by the previous work.  In fact, almost any amount of hard-band flux
could be obtained if there are further even more opaque partial covering layers or 
heavily-absorbed reflection zones, and recently the source PDS\,456 has been
found to exhibit just such a heavily absorbed zone (Reeves et al. in preparation).
If the narrow Fe\,K$\alpha$ emission line originates in optically thin gas rather
than optically-thick reflection, it may even be the case that the heavily
absorbed reflection component could be largely replaced by further high-opacity
partial covering layers.  We do not explore such models further
in this paper.

\subsection{The location of the Fe emission-line region}
A number of previous authors have suggested using the width of the narrow
6.4\,keV\,Fe\,K$\alpha$ emission line to give an indication of its origin
\citep{lee02,yaqoob04}.  
In addition to the red wing, there appears to be a resolved
6.4\,keV\,Fe\,K$\alpha$ line whose width, if interpreted as due to Doppler
broadening, is FWHM$\sim 10,000$\,km\,s$^{-1}$.  In the model presented
here, some of this width is provided by a Compton shoulder on the line in a
moderately ionised reflector, modelled by {\sc reflionx}, although there may
still be some residual excess emission on the blue side of the line. In this
case it is difficult to disentangle Compton broadening from velocity
broadening and the line width gives no clear indication of the location of
the reflection, other than not being close to the black hole.
If the line instead is emitted from optically-thin gas, the
linewidth may indicate an origin in the broad line region.  In addition to this
component, the long \chandra exposure reveals the presence of a weak
(equivalent width $\sim 20$\,eV) line unresolved at the {\sc heg} resolution
(FWHM $< 3000$\,km\,s$^{-1}$) that may indicate a further component of reflection
or emission even more distant from the central source.

\subsection{The location of the absorbers, rapid variability and time delays}
The analysis presented here has only dealt with variability on timescales
$\ga 20$\,ks.  Within the model presented, there is no requirement for any
material within radii where relativistic blurring is significant ($\la 20$\,r$_g$) 
but we suggest that the heavy partial-covering absorbing layer originates
on scales of the accretion disc, perhaps around 100\,r$_g$.  Perhaps the most
compelling evidence for this is the observation of an apparent eclipse-like event
in the light curve of MCG$-6$-30-15 \citep{mckernan98} which may be explained
as a clumpy disc wind occulting part of the source.  M07 and T07 also
suggested absorption by a clumpy disc wind in Mrk\,766 and a similar eclipse
event to that seen in MCG$-6$-30-15 has been observed in NGC\,3516 \citep{turner08}.
Full spectral models of such a wind do not currently exist, although 
\citet{schurch07} have created approximate spectra by layering (1D) {\sc xstar}
absorption zones and \citet{sim08}
are developing steady axisymmetric wind models based on Monte-Carlo radiative transfer 
that qualitatively reproduce many of the features seen in AGN and Galactic
black hole binary system X-ray spectra.  \citet{dorodnitsyn08} have calculated
transmission spectra in a time-dependent axisymmetric model of parsec-scale
flows in AGN. We can hope that detailed comparison of observed spectra with
realistic wind models will become possible in the near future.

In the model presented here there is also a component of low-ionisation reflection.
the amplitude of this is so constant that it is very likely from distant material,
light-days or further away,
with any reflection variability erased by light travel time delays.  However,
\citet{ponti04} have found evidence for a transient reflection signal delayed
after a continuum flare by a few ks.  The statistical significance of the 
delayed flare is low and needs to be confirmed by detection of further similar
events.  
\citet{goosmann07} have modelled the phenomenon as reflection from a dense clumpy medium
surrounding the accretion region at $\sim 70$\,r$_g$.  
If our hypothesised absorbing clumpy disk wind is present, we might
expect to see some reflection contribution from it, depending on its scattering
optical depth and covering factor.
We suggest that the clumpy disc wind 
may also be visible in reflection on short-lived
occasions after a flare when viewed with adequate time resolution.  On the
20\,ks timescales used in the analysis in this paper such reflection,
weak in normal circumstances, would simply be included as an additional component
in the variable component of the spectrum (PCA eigenvector one).

We may also be able to constrain the location of the absorbing zones from their
response, or lack of, to the continuum variations.  Of the various zones, only
the high ionisation outflow, zone 3, seems to show evidence for ionisation
variation that tracks the continuum brightness (section\,\ref{ionvar}).
This interpretation of the change in equivalent width is not unique, it could
also arise if zone 3 were only associated with the heavily-absorbed
components, but if the ionisation-variation interpretation is correct it implies
that the recombination time in the absorbing zone 
is no larger than the typical continuum variability timescale, 
and in turn that the depth
of the zone should be $\la 10$ light days, 
assuming a \ion{Fe}{xxvi} radiative recombination rate coefficient
$\alpha \simeq 10^{-11}$\,cm$^3$\,s$^{-1}$ \citep{shull}.
This might place the
wind within the broad-line region.  In the other absorbing zones that are
detected in the grating data, there do also appear to be ionisation variations,
but these do not appear to be systematically correlated with the source brightness
\citep{gibson07}, which might imply that their densities are rather low, or else
that more complex radiative transfer is present.  There is currently no constraint
on ionisation variation of the heavily absorbing zones 4 and 5.

\subsection{Comparison with other AGN}
Much of the analysis in this paper has followed that carried out for Mrk\,766,
a narrow-line Seyfert\,I, by M07 and T07.  There are strong similarities between
the X-ray spectra of the two sources:
\begin{enumerate}
\setlength\itemsep{0em}
\item Both sources show similar systematic variation in spectral shape with flux
on 20\,ks timescales, with a hard component in the low flux state that shows a 
significant edge around the Fe K$\alpha$ region.  
\item The PCA leads to similar spectral components in both sources, with the
``red wing'' component being dominated by an apparently quasi-constant
hard spectral component.
\item Both sources appear to have absorption from a high-ionisation
outflow, which leads to significant modification of the observed spectral shape
around 7\,keV.
\item The soft excess appears to be explained by ionised absorption.
\end{enumerate}
The implication is that MCG$-6$-30-15 is not a special case, but that it is a
examplar of a general class of AGN whose X-ray spectra are dominated by the
effects of absorption, possibly in an outflowing wind. \citet{nandra07} have
attempted to quantify the occurrence of ``red wings'' in AGN, with the aim of
establishing the prevalence of detectable relativistic blurring. They find 
that 45\,percent of AGN in their sample are best fit by a relativistically
blurred component, including MCG$-6$-30-15 and Mrk\,766.  The results we obtain
suggest that a partial covering model such as presented here would provide an
alternative explanation of the observed red wings in AGN. The high occurrence
rate in the \citet{nandra07} sample would then imply a high prevalence of
significant wind absorption, that in turn would imply a high global covering
fraction for the wind.  
If the wind explanation is correct, 
the inference of X-ray winds in the detailed studies of three AGN, 
Mrk\,766, MCG$-6$-30-15 and NGC\,3516 \citep{turner08},
would require that the phenomenon be a property of sources across the range
of narrow-line Seyfert\,I and broad-line Seyfert\,I AGN characteristics,
across a range of black hole mass, as indicated by the
wide range of sources for which a red wing 
is claimed in the \citeauthor{nandra07} study. 

\section{Conclusions} 
We have investigated a model of X-ray spectral variability for MCG$-6$-30-15
based on the absorbing zones identified in high-resolution grating data.
We find the ``soft excess'' may be explained entirely by the combined effect of those
zones (including soft-band dust absorption).  
High-resolution principal components analysis, achieved using singular value
decomposition,  indicates the presence of a less variable heavily absorbed
component that until now has been interpreted as a relativistically-blurred
Fe line.  This component may be modelled by a combination of distant (constant
amplitude) absorbed reflection and the effect of a variable covering fraction
of absorption of the primary continuum source.  
The model has been applied 
both to the PCA and the actual data accumulated from the
\xmm {\sc pn} and {\sc rgs} instruments (simultaneously-fitted)
in 2000 and 2001 over the energy range
$0.5-10$\,keV, to the \suzaku {\sc xis}, $0.5-10.5$\,keV, and 
{\sc pin}, $15-45$\,keV, data (simultaneously fitted) from 2006 
and to the \chandra {\sc hetgs} data 
({\sc heg} and {\sc meg} simultaneously fitted) from 2004.  
This is the most
comprehensive analysis of the MCG$-6$-30-15 dataset yet published.
Remarkably, the absorption model fits the entire dataset over its entire
range, explaining simultaneously the soft-band excess, the ``red wing'' and
its lack of variability and the high hard band 
({\em BeppoSAX} and \suzaku {\sc pin}) flux and its lack of variability,
and fits not only the CCD-resolution data but also matches well the
absorption lines and edges seen in the high resolution grating data.
The best-fit parameters show that the partial-covering absorber 
is ionised, but with $\xi < 100$\,erg\,cm\,s$^{-1}$,
so no Fe\,K$\alpha$ absorption is expected from this component, 
and the absence of observed 6.5\,keV\,Fe\,K$\alpha$ absorption \citep{young05}
is not therefore a constraint on this model.
No relativistically blurred component is required to fit this dataset.
We suggest the absorbing material is primarily a clumpy disc wind.

\begin{acknowledgements}
This paper is based 
on observations obtained with {\em XMM-Newton}, \suzaku and {\em Chandra}.
\xmm is an ESA science mission with instruments and contributions 
directly funded by ESA Member States and NASA.
\suzaku is a collaboration between 
ISAS/JAXA, NASA/GSFC and MIT. 
This research has made use of data obtained from the High Energy Astrophysics 
Science Archive Research Center (HEASARC), provided by NASA's 
Goddard Space Flight Center.
TJT acknowledges NASA grant ADP03-0000-00006. 
\end{acknowledgements}

\label{lastpage}

\end{document}